\documentclass[prb,aps, 10pt,twocolumn,superscriptaddress,floatfix,notitlepage,citeautoscript]{revtex4-2}
\usepackage[pdftex]{graphicx}
\usepackage{amsmath}
\usepackage{color}
\usepackage{hyperref}
\usepackage[usenames,dvipsnames]{xcolor}
\hypersetup{colorlinks=true,linkcolor=blue,urlcolor=blue,citecolor=blue}

\begin{document}
\title{Coupled system of electrons and exciton-polaritons: \\ Screening, dynamical effects, and superconductivity}

\author{Andrey S. Plyashechnik}%
\affiliation{National Research Nuclear University MEPhI (Moscow Engineering Physics Institute), 115409 Moscow, Russia}%
\author{Alexey A. Sokolik}%
\affiliation{Institute for Spectroscopy RAS, 142190 Troitsk, Moscow, Russia}%
\affiliation{National Research University Higher School of Economics, 109028 Moscow, Russia}%
\author{Nina S. Voronova}%
\email{nsvoronova@mephi.ru}%
\affiliation{National Research Nuclear University MEPhI (Moscow Engineering Physics Institute), 115409 Moscow, Russia}%
\affiliation{Russian Quantum Center, Skolkovo Innovation Center, 121205 Moscow, Russia}%
\author{Yurii~E. Lozovik}%
\affiliation{Institute for Spectroscopy RAS, 142190 Troitsk, Moscow, Russia}%
\affiliation{National Research University Higher School of Economics, 109028 Moscow, Russia}%

\begin{abstract}
Bose-Fermi systems such as mixtures of electrons with excitons or exciton-polaritons are extensively discussed as candidates to host a variety of intriguing phenomena, including polaron formation, drag effects, supersolidity, and superconductivity.
In this work, assuming the strong-coupling regime between the semiconductor excitons and cavity photons,
we develop the many-body theory approach addressing the interplay of different types of interaction among various species in such a mixture, wherein we take into account dynamical density responses of both the Bose-condensed exciton-polaritons and the two-dimensional electron gas inside an optical microcavity.
As was anticipated previously, at high enough polariton densities the lower hybrid mode of the system's excitation spectrum acquires a roton minimum, making the system prone to superconducting pairing in the vicinity of the roton instability.
We analyze the possibility of
polariton-BEC-mediated superconductivity in the electron gas taking into account full momentum and frequency dependence of the gap, as well as in the Eliashberg approach where the momentum dependence is neglected, and in the Bardeen-Cooper-Schrieffer approach that discards the  frequency dependence and dynamical effects. Considering the interaction screening in Thomas-Fermi and in random-phase approximations, we estimate the critical temperatures of superconductivity to be not larger than 0.1~K in the vicinity of instability. As possible realizations of the coupled polariton-electron system, semiconductor quantum wells and two-dimensional transition metal dichalcogenides are considered.
\end{abstract}

\maketitle

\section{Introduction}\label{Sec1}

In the recent years, with the advent of two-dimensional transition metal dichalcogenides (TMDs) hosting strongly-bound and polarization-controlled excitons, trions, biexcitons, and other complexes, the many-body physics of Bose-Fermi mixtures of excitons and electrons is intensively shifting from a theoretical possibility to an experimentally-approachable challenge.
Placing TMDs into an optical microcavity allows to couple excitonic complexes with light \cite{Wang,Berkelbach,Vincent,Huang,Lundt,Dufferwiel,Ardizzone,Rana}, while stacking TMDs into Moir\'{e} heterostructures \cite{moire1,Mak,Shimazaki,Kennes} opens new degrees of freedom to manipulate both the Bose and Fermi quasiparticles in such systems.

Exciton-polaritons resulting from light-matter coupling in optical microcavities were extensively studied in a variety of material systems based on quantum wells~\cite{RMP2010}, molecular dyes~\cite{Lidzey,Lagoudakis}, perovskites~\cite{Fieramosca,Su}, and TMDs~\cite{Lundt,Dufferwiel}. Their most striking feature is the ability to form macroscopically-coherent states like Bose-Einstein condensates (BECs)~\cite{Byrnes,Su,Schneider} and superfluid phases~\cite{QFL}.
Until recently, however, all related phenomena were limited by cryogenic temperatures and generally reduced exciton-exciton interactions. In the last decade, with the emergence of organic materials and TMD-based heterostructures, the situation rapidly changes as the reduced dimensionality and large binding energies bring excitons into the regime of strong interactions~\cite{Rapaport,Fieramosca,Xu} and stability at room temperature~\cite{Ardizzone}.

When a two-dimensional electron gas (2DEG) is added as an additional layer, among the effects predicted for strongly coupled electron-exciton or electron-polariton systems due to interaction between different species are the Coulomb and superfluid drag~\cite{Berman,Boev,Sokolik}, the roton instability formation~\cite{Shelykh,Shelykh2} and exciton supersolidity~\cite{supersolid}. Another frontier of Bose-Fermi systems is the polaron-polariton physics emerging in TMDs when the electron density is increased \cite{Imamoglu1,Bastarrachea-Magnani,Zhumagulov,Imamoglu2,Muir,Julku2021}. If Bose condensation occurs in the exciton or exciton-polariton subsystem, the many-body processes between the BEC and the 2DEG may become enhanced due to the effect of Bose stimulation. One of such new phenomena theoretically proposed is the superconductivity mediated by an exciton or polariton BEC \cite{Laussy_JNP,Cherotchenko,Sedov,Sun_2DMater,Sun_Thesis,Sun_NJP,Sun_PRR,Skopelitis,Laussy_PRL,Cotlet,Kinnunen,Julku}. Unlike the exciton mechanism of superconductivity proposed by W.A.~Little \cite{Little}, where the pairing interaction occurs due to virtual excitons, here it is mediated by virtual Bogoliubov excitations (or bogolons) emerging from the exciton or polariton BEC. This setup has two advantages. First is the stimulation of bogolon creation and annihilation in the presence of a BEC, which enhances the pairing interaction proportionally to the number of Bose-condensed particles. Second, thanks to their gapless dispersion and generally lower energies, the Bogoliubov modes provide enhanced contribution to the resulting coupling constant in comparison to excitons. The polariton-BEC setting is especially attractive, compared to the exciton BEC, due to high critical temperature of the BEC formation \cite{Byrnes,Schneider}.

The aim of this paper is to develop a consistent many-body description for the system of Bose-condensed exciton-polaritons
strongly coupled to a 2DEG, taking into account interaction screening and dynamical effects. Previous works on such systems assumed unscreened \cite{Laussy_JNP,Cherotchenko,Sedov,Sun_2DMater,Sun_Thesis,Sun_NJP,Sun_PRR,Skopelitis} or  only statically screened \cite{Laussy_PRL,Cotlet} interactions and overlooked effects of dynamical screening. Therefore, the aim of this work is to revisit the potentiality of superconducting pairing in the 2DEG mediated by virtual Bogoliubov excitations of the polariton BEC. According to our estimates, contrary to the previous predictions~\cite{Laussy_PRL,Laussy_JNP,Cherotchenko,Sedov,Sun_2DMater,Sun_Thesis,Sun_NJP,Sun_PRR} (while in agreement with \cite{Cotlet}), the exciton-polariton mechanism of superconductivity in the 2DEG can be realized only in a very narrow vicinity of the roton instability because of the dominating contribution of low-frequency roton excitations to the pairing. We also show that the pair-bogolon processes which were claimed to dominate the pairing \cite{Sun_2DMater,Sun_Thesis,Sun_NJP,Sun_PRR} in fact provide negligible contribution when the interaction screening is taken into account. Our calculations rely on the mean-field approaches that are widely used in the many-body analysis of electronic and polaritonic systems, such as the Bogoliubov theory, Gor'kov equations, and random-phase approximation for interaction screening. Although some non-perturbative correlation effects may be overlooked in these descriptions, especially in a close vicinity of the roton instability, the main goal of our work is to reveal the important role of screening and dynamical effects, in particular in the context of polariton-BEC-mediated superconductivity, even at the mean-field theory level.

\begin{figure}[b]
\begin{center}
\includegraphics[width=0.6\columnwidth]{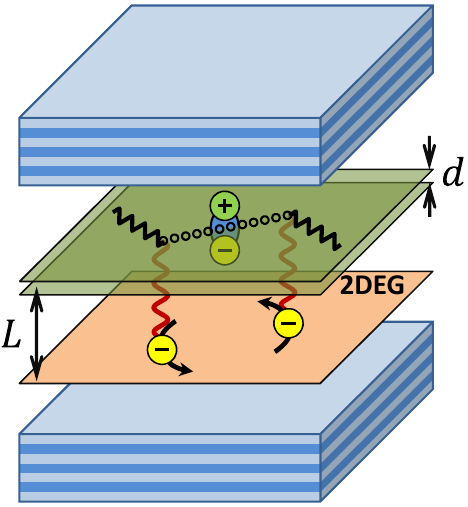}
\end{center}
\caption{\label{Fig_Scheme}System schematic: semiconducting (TMD or QW) bilayer with the interlayer separation $d$ hosting dipolar excitons at the distance $L$ from the 2DEG layer inside Fabry-P\'{e}rot optical microcavity. Exchange by virtual Bogoliubov excitations (dotted line) knocked out from the polariton BEC (zigzag lines) induces effective attraction between electrons in the 2DEG.}
\end{figure}

The paper is organized as follows. In Sec.~\ref{Sec2} we introduce the system, describe the screening of interactions by density responses of both the exciton and 2DEG layers, and provide the parameters that we use throughout the paper. For the 2DEG density response, we consider both the static Thomas-Fermi and dynamical random-phase approximations (TFA and RPA, respectively). In Sec.~\ref{Sec3} we introduce the spectrum of excitations consisting of two hybrid modes that originate from the coupling of Bogoliubov quasiparticles on top of the polariton BEC with plasmons and single-particle excitations of the 2DEG. We resolve the contributions of different excitation branches to the electron-electron pairing interaction by characterizing them using Eliashberg functions and dimensionless coupling constants. Many-body theory of superconducting pairing is described in Sec.~\ref{Sec4} where we employ and compare several widely used approaches. The most elaborate analysis takes into account both the momentum and frequency dependence of the gap and renormalization functions. The Eliashberg approach (and the analytical Allen-Dynes formula approximating its results) catches only the dynamical effects using frequency-dependent functions, while the momentum dependence is neglected. The Bardeen-Cooper-Schrieffer (BCS) approach includes only the momentum dependence and disregards dynamical effects. We revisit the problem of polariton-BEC-mediated superconducting pairing in all three approaches and demonstrate the importance of interaction screening and dynamical effects, showing that the BCS approach provides greatly overestimated critical temperatures of superconductivity. Our conclusions are stated in Sec.~\ref{Sec5}. Appendix~\ref{Appendix_A} contains simple formulae which can be used to estimate the superconducting coupling constant and critical temperature, Appendix~\ref{Appendix_B} provides details of the momentum and frequency dependence of the gap and renormalization functions, Appendix~\ref{Appendix_C} demonstrates the difference between the Eliashberg and BCS approaches, and Appendix~\ref{Appendix_D} is devoted to analysis of the role of pair-bogolon processes.

\section{Electron-polariton system}\label{Sec2}

The schematic of the system is shown in Fig.~\ref{Fig_Scheme}. We assume that the exciton layer hosts dipolar excitons
coupled to microcavity photons to form polaritons. Possible material realizations are based on recently emerging TMD bilayers \cite{gerber,lorchat,menon,tartakovskii} or, alternatively, well-studied conventional semiconductor coupled quantum wells (QW). Coupled electron-polariton system is described by the Hamiltonian
\begin{equation}
H=H_\mathrm{e}+H_\mathrm{p}+H_\mathrm{e-e}+H_\mathrm{p-p}+H_\mathrm{e-p},
\end{equation}
where $H_\mathrm{e}=\sum_{\mathbf{k}s}\epsilon^\mathrm{e}_kc^+_{\mathbf{k}s}c_{\mathbf{k}s}$, $H_\mathrm{p}=\sum_\mathbf{k}\epsilon_k^\mathrm{p}b^+_\mathbf{k}b_\mathbf{k}$
are the Hamiltonians of bare electrons and lower polaritons whose annihilation operators are, respectively, $c_{\mathbf{k}s}$ and $b_\mathbf{k}$ at momentum $\mathbf{k}$ ($s$ is the electron spin and valley index, and the polariton spin is omitted). The dispersion relation for electrons with the effective mass $m_\mathrm{e}^*$ in the 2DEG is $\epsilon^\mathrm{e}_k=k^2/2m_\mathrm{e}^*$ (hereafter we assume $\hbar\equiv1$), and for the lower polaritons we take the standard dispersion provided by the coupled-oscillator model (see e.g. \cite{Cotlet}):
\begin{equation}
\epsilon^\mathrm{p}_k=\frac12\left\{\sqrt{\delta^2+4\Omega_\mathrm{R}^2}+\frac{k^2}{2m_\mathrm{p}}-\sqrt{(\delta E_k)^2+4\Omega_\mathrm{R}^2}\right\}.\label{e_p}
\end{equation}
Here $m_\mathrm{p}^{-1}=m_\mathrm{x}^{-1}+m_\mathrm{c}^{-1}$, with $m_\mathrm{x}$ and $m_\mathrm{c}$ being the exciton and cavity photon effective masses, respectively, $\Omega_\mathrm{R}$ is the Rabi frequency, $\delta E_k=\delta+k^2(m_\mathrm{c}^{-1}-m_\mathrm{x}^{-1})/2$ is the difference between photon and exciton energy dispersions, $\delta$ is the photon-to-exciton detuning at zero momentum.

Bare electron-electron Coulomb interaction is
\begin{equation}
H_\mathrm{e-e}=\frac1{2S}\sum_{\mathbf{k}\mathbf{k}'\mathbf{q}}\sum_{ss'}v_q^\mathrm{ee}c^+_{\mathbf{k}+\mathbf{q},s}c^+_{\mathbf{k}'-\mathbf{q},s'}c_{\mathbf{k}'s'}c_{\mathbf{k}s},
\end{equation}
with $v_q^\mathrm{ee}=2\pi e^2/\varepsilon q$ being the Fourier image of the Coulomb potential screened by the dielectric constant $\varepsilon$ of the surrounding medium. Polariton-polariton interaction
\begin{equation}
H_\mathrm{p-p}\!=\!\frac1{2S}\!\!\sum_{\mathbf{k}\mathbf{k}'\mathbf{q}}\!v^\mathrm{xx}_{q}\!X_{|\mathbf{k}+\mathbf{q}|}X_{|\mathbf{k}'\!-\mathbf{q}|}X_{k'}X_k b^+_{\mathbf{k}+\mathbf{q}}b^+_{\mathbf{k}'\!-\mathbf{q}}b_{\mathbf{k}'}b_\mathbf{k}\label{x-x}
\end{equation}
here for simplicity is assumed to be governed by the contact exciton-exciton interaction $v^\mathrm{xx}_{q}=U$ dressed with the Hopfield coefficients
\begin{equation}
X_k=\sqrt{\frac12\left\{1+\frac{\delta E_{k}}{\sqrt{(\delta E_k)^2+4\Omega_\mathrm{R}^2}}\right\}}
\end{equation}
which determine the amplitude of the exciton component of the lower polariton wave function. Strictly speaking, dipolar exciton-polaritons that we consider here assume the dispersion law differing from the regular lower-polariton dispersion (\ref{e_p}) since dipolaritons in TMD bilayers (or in QW) are mixtures of three modes: photon, direct (intralayer) and indirect (interlayer) excitons~\cite{Baumberg,Yamamoto,menon,tartakovskii}. However, as discussed in~\cite{Yamamoto}, the BEC of dipolaritons occurs on the lowest of the three branches and hence the deviation of their bare-particle dispersion from (\ref{e_p}) is negligible. Moreover, the interaction matrix element of indirect excitons in bilayers may strongly depend on transfer momentum~\cite{Maslova}, thus the assumption of contact interaction made in (\ref{x-x}) is not always applicable. Nevertheless, as will be discussed below, only $v^\mathrm{xx}_{q}$ at $q$ near the roton minimum is important for such many-body effects as roton instability and superconductivity, so the full dependence of $v^\mathrm{xx}_{q}$ on $q$ in a wide range of momenta may be omitted.

Finally, the electron-polariton interaction
\begin{equation}
H_\mathrm{e-p}=\frac1S\sum_{\mathbf{k}\mathbf{k}'\mathbf{q}s}v^\mathrm{ex}_{q}X_{|\mathbf{k}+\mathbf{q}|}X_kb^+_{\mathbf{k}+\mathbf{q}}c^+_{\mathbf{k}'-\mathbf{q},s}c_{\mathbf{k}'s}b_{\mathbf{k}}
\end{equation}
is defined by the Fourier image $v_q^\mathrm{ex}$ of the electron-exciton interaction dressed with the Hopfield coefficients. For the former, we take the following form accounting for the dipole moment of indirect excitons and in-plane spread of their wave function \cite{Laussy_JNP,Cotlet}:
\begin{equation}\label{v_ex}
v_q^\mathrm{ex}=\frac{2\pi e^2}{\varepsilon q}\left\{\frac{e^{-q(L-\beta_\mathrm{e}d)}}{[1+(\frac{\beta_\mathrm{e}qa_\mathrm{B}}2)^2]^{\frac32}}-\frac{e^{-q(L+\beta_\mathrm{h}d)}}{[1+(\frac{\beta_\mathrm{h}qa_\mathrm{B}}2)^2]^{\frac32}}\right\}.
\end{equation}
Here $\beta_\mathrm{e,h}=m_\mathrm{e,h}/m_\mathrm{x}$, and $m_\mathrm{x}=m_\mathrm{e}+m_\mathrm{h}$, $m_\mathrm{e,h}$ are the effective masses of an electron and a hole making up the exciton, $d$ is the interlayer separation between electrons and holes in the bilayer hosting excitons, $L$ is the out-of-plane distance between the centers of masses of the electron and exciton wave functions (see Fig.~\ref{Fig_Scheme}), and $a_\mathrm{B}$ is the exciton in-plane Bohr radius.

\begin{figure}
\begin{center}
\includegraphics[width=\columnwidth]{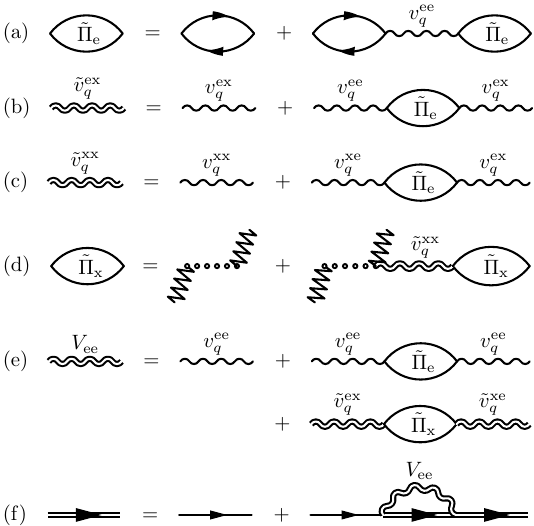}
\end{center}
\caption{\label{Fig_Diag}Feynman diagrams describing our approach: (a) dressing (\ref{Pi_e_scr}) of the 2DEG density response by the Coulomb interaction; (b,c) renormalization of the electron-exciton (\ref{v_ex_tilde}) and exciton-exciton (\ref{tilde_v_xx}) interactions, respectively, by the density response of the 2DEG (here the symmetry $v_q^\mathrm{ex}=v_q^\mathrm{xe}$, $\tilde{v}_q^\mathrm{ex}=\tilde{v}_q^\mathrm{xe}$ is implied); (d) dressing (\ref{tilde_Pi_x}) of the excitonic density response by interactions; (e) screened electron-electron interaction as a sum (\ref{V_ee_decomp}) of intralayer interaction in the 2DEG ($V_1$, first line of diagrams) and polariton-induced interaction ($V_2$, second line of diagrams); (f) Gor'kov equations (\ref{Gorkov_eq}) for the electron Cooper pairing. In all panels, the straight single and double lines represent the bare and dressed electron Green's functions, respectively; single and double wiggled lines represent the bare and screened interactions among different kinds of particles; the dotted line is the bare polariton Green's function, and the zigzag lines denote the BEC of polaritons.}
\end{figure}

The many-body approximations we use are summarized in terms of Feynman diagrams in Fig.~\ref{Fig_Diag}. The main quantity of interest in our approach is the screened electron-electron interaction $V_\mathrm{ee}$ which contains both the direct Coulomb repulsion of electrons in an isolated 2DEG and effective attraction caused by virtual Bogoliubov excitations in the neighboring polariton system. It is given by the standard electrodynamic formula for the screened intralayer interaction in a double-layer system:
\begin{equation}
V_\mathrm{ee}(q,i\omega_n) = \frac{v_q^\mathrm{ee}-v_q^\mathrm{ee}v^\mathrm{xx}_q\Pi_\mathrm{x}+(v^\mathrm{ex}_q)^2\Pi_\mathrm{x}}{(1 \!-\! v_q^\mathrm{ee}\Pi_\mathrm{e})(1 \!-\! v^\mathrm{xx}_q\Pi_\mathrm{x}) \!-\!(v^\mathrm{ex}_q)^2\Pi_\mathrm{e}\Pi_\mathrm{x}}.
\label{V_ee}
\end{equation}
Here $\Pi_\mathrm{e}$ is the irreducible density response function of the 2DEG,  $\Pi_\mathrm{x}$ is the density response of excitons in the polariton system. Both $\Pi_\mathrm{e,x}(q,i\omega_n)$ are considered at a given wave vector $q$ and bosonic Matsubara frequency $i\omega_n=2\pi i n T$, where $T$ denotes the temperature. We take $\Pi_\mathrm{e}$ either in RPA where it is given by the density response function of a noninteracting 2DEG \cite{Stern}, or in TFA which is applicable at $q\rightarrow0$, $\omega=0$ and yields $\Pi_\mathrm{e}^\mathrm{TFA}=-gm_\mathrm{e}^*/2\pi$. Here $g$ is the spin or spin-valley degeneracy factor which is equal to 2 for QW and 4 for TMD. Note that TFA was
used in Refs.~\cite{Laussy_JNP,Cherotchenko} studying the polariton mechanism of superconductivity and in Ref.~\cite{Boev} studying electron-polariton drag.

We assume the polariton system in the BEC state at the temperatures much lower than the critical temperature $T_\mathrm{BEC}$ of Bose-Einstein condensation. At these conditions, the polariton system can be described by the Bogoliubov theory \cite{Shi} implying that vast majority of polaritons belong to the condensate, so that the condensate density $n_0^\mathrm{p}$ is almost equal to the total density of polaritons $n_\mathrm{p}$. The excitonic density response $\Pi_\mathrm{x}$ of the polariton system in RPA is a response of noninteracting system of polaritons. At $T\ll T_\mathrm{BEC}$ it is dominated by the condensate processes \cite{Griffin} with excitation of a single particle out of the condensate:
\begin{equation}
\Pi_\mathrm{x}(q,i\omega_n)=\frac{2X_0^2X_q^2n_0^\mathrm{p}\tilde\epsilon_q^\mathrm{p}}{(i\omega_n)^2-(\tilde\epsilon_q^\mathrm{p})^2}.\label{Pi_x}
\end{equation}
The role of the Hopfield coefficients $X_{0,q}$ here is to relate the polariton density to the excitonic one. We note that the polariton dispersion (\ref{e_p}) is renormalized here by the Hartree mean-field interaction with the condensate~\cite{Grudinina}: $\tilde\epsilon_q^\mathrm{p}=\epsilon_q^\mathrm{p}+X_0^2(X_q^2-X_0^2)n_0^\mathrm{p}v_0^\mathrm{xx}$.

It is useful to introduce the density response of the interacting electron gas [the corresponding diagram is shown in Fig.~\ref{Fig_Diag}(a)] as
\begin{align}
\tilde\Pi_\mathrm{e}(q,i\omega_n) & = \Pi_\mathrm{e}(q,i\omega_n)+\Pi_\mathrm{e}(q,i\omega_n)v_q^\mathrm{ee}\tilde\Pi_\mathrm{e}(q,i\omega_n)\nonumber\\
& = \frac{\Pi_\mathrm{e}(q,i\omega_n)}{1-v_q^\mathrm{ee}\Pi_\mathrm{e}(q,i\omega_n)},\label{Pi_e_scr}
\end{align}
which renormalizes both the electron-exciton [Fig.~\ref{Fig_Diag}(b)]
\begin{align}
\tilde{v}^\mathrm{ex}_q(i\omega_n)&=v^\mathrm{ex}_q+
v^\mathrm{ee}_q\tilde\Pi_\mathrm{e}(q,i\omega_n)v^\mathrm{ex}_q\nonumber\\
&=\frac{v^\mathrm{ex}_q}{1-v_q^\mathrm{ee}\Pi_\mathrm{e}(q,i\omega_n)}\label{v_ex_tilde}
\end{align}
and the exciton-exciton [Fig.~\ref{Fig_Diag}(c)]
\begin{align}
\tilde{v}_q^\mathrm{xx}(i\omega_n)&=v_q^\mathrm{xx}+v_q^\mathrm{xe}\tilde\Pi_\mathrm{e}(q,i\omega_n)v_q^\mathrm{ex}\nonumber\\&=v_q^\mathrm{xx}+(v^\mathrm{ex}_q)^2\frac{\Pi_\mathrm{e}(q,i\omega_n)}{1-v_q^\mathrm{ee}\Pi_\mathrm{e}(q,i\omega_n)}\label{tilde_v_xx}
\end{align}
interactions. Physically, the second term in (\ref{tilde_v_xx}) describes the contribution of virtual electron-hole pair and plasmon excitations in the 2DEG to the exciton-exciton interaction.

Bose-condensed system of polaritons interacting via (\ref{tilde_v_xx}) is characterized by the Bogoliubov self-energies
\begin{equation}  E_q^\mathrm{p}(i\omega_n)=\sqrt{\tilde\epsilon_q^\mathrm{p}[\tilde\epsilon_q^\mathrm{p}+2X_0^2X_q^2n_0^\mathrm{p}\tilde{v}_q^\mathrm{xx}(i\omega_n)]}.\label{Bog_disp}
\end{equation}
Since the interaction $\tilde{v}_q^\mathrm{xx}(i\omega_n)$ is generally frequency-dependent due to the dynamical screening, $E_q^\mathrm{p}(i\omega_n)$ depends on frequency, hence the bogolon dispersion $\omega(q)$ should be found self-consistently as $\omega(q)=E_q^\mathrm{p}(\omega(q))$. The excitonic density response of the interacting polariton system [shown in the diagrammatic form in Fig.~\ref{Fig_Diag}(d)]
\begin{multline}
\tilde\Pi_\mathrm{x}(q,i\omega_n)=\Pi_\mathrm{x}(q,i\omega_n)+\Pi_\mathrm{x}(q,i\omega_n)\tilde{v}_q^\mathrm{xx}(i\omega_n)\tilde\Pi_\mathrm{x}(q,i\omega_n)\\
\!=\! \frac{\Pi_\mathrm{x}(q,i\omega_n)}{1 \!-\! \tilde{v}_q^\mathrm{xx}(i\omega_n)\Pi_\mathrm{x}(q,i\omega_n)}
\!=\! \frac{2X_0^2X_q^2n_0^\mathrm{p}\tilde\epsilon_q^\mathrm{p}}{(i\omega_n)^2 \!-\! [E_q^\mathrm{p}(i\omega_n)]^2}
\label{tilde_Pi_x}
\end{multline}
has poles at the bogolon energies (\ref{Bog_disp}).

Using the notations (\ref{v_ex_tilde}) and (\ref{tilde_Pi_x}) in (\ref{V_ee}), we can represent the screened electron-electron interaction as [Fig.~\ref{Fig_Diag}(e)]
\begin{align}
V_\mathrm{ee}(q,i\omega_n)&=V_1(q,i\omega_n)+V_2(q,i\omega_n),\label{V_ee_decomp}\\
V_1(q,i\omega_n)&=\frac{v_q^\mathrm{ee}}{1-v_q^\mathrm{ee}\Pi_\mathrm{e}(q,i\omega_n)},\label{V_1}
\\
V_2(q,i\omega_n)&=(\tilde{v}^\mathrm{ex}_q)^2\tilde\Pi_\mathrm{x}(q,i\omega_n).\label{V_2}
\end{align}
Here $V_1$ is the dynamically screened interaction in the 2DEG, which exists in the absence of polaritons, and $V_2$ is the polariton-induced contribution to the electron-electron interaction mediated by virtual Bogoliubov excitations ($\tilde\Pi_\mathrm{x}$) in the polariton system through the second-order interlayer interaction ($\tilde{v}^\mathrm{ex}_q$) screened by the 2DEG.

In our calculations in the following sections, we consider two possible experimental setups, based on TMD bilayers and coupled QWs embedded into an optical microcavity. In the TMD-based setup, we consider dipolar excitons with the out-of-plane electron-hole distance $d=1\,\mbox{nm}$ \cite{gerber,lorchat,leisgang}. We take electron and hole effective masses $m_\mathrm{e}\approx m_\mathrm{h}= 0.5m_0$ (where $m_0$ is the free electron mass) and the dielectric constant of the surrounding medium $\varepsilon=4.4$, which are the common parameters of TMD bilayers encapsulated in hexagonal boron nitride, see e.g.~\cite{Komsa,Laturia,Goryca}. The corresponding exciton Bohr radius is of the order of $a_\mathrm{B}=1\,\mbox{nm}$. For the 2DEG layer that we consider being a doped TMD monolayer separated from the exciton center-of-mass by the distance $L=2\,\mbox{nm}$ (see Fig.~\ref{Fig_Scheme}), we take the electron effective mass $m_\mathrm{e}^*=0.5m_0$ and
density $n_\mathrm{e}=10^{13}\,\mbox{cm}^{-2}$. The photon effective mass in the microcavity is $m_\mathrm{c}=4\times10^{-5}m_0$ and the Rabi splitting is $\Omega_\mathrm{R}=20\,\mbox{meV}$ \cite{Schneider,menon,tartakovskii}. To enhance electron-exciton interaction by increasing the excitonic fraction in a lower polariton, we assume positive photon-to-exciton detunings $\delta=40\,\mbox{meV}$. The density of polaritons $n_\mathrm{p}=7.8987\times10^{12}\,\mbox{cm}^{-2}$ is taken
slightly below the critical density for the roton instability $n_\mathrm{p}^\mathrm{crit}=7.8989\times10^{12}\,\mbox{cm}^{-2}$.

For the QW-based setup, we consider three-layered system with the typical GaAs parameters: $d=9\,\mbox{nm}$, $m_\mathrm{e}=m_\mathrm{e}^*=0.067m_0$, $m_\mathrm{h}=0.45m_0$, $\varepsilon=13$, $a_\mathrm{B}=7\,\mbox{nm}$, $n_\mathrm{e}=10^{12}\,\mbox{cm}^{-2}$, $L=13.5\,\mbox{nm}$. The microcavity parameters in this case are $m_\mathrm{c}=4\times10^{-5}m_0$, $\Omega_\mathrm{R}=5\,\mbox{meV}$, $\delta=10\,\mbox{meV}$ (see e.g. \cite{RMP2010,Byrnes,QFL} and references therein). To illustrate the role of the roton minimum and study the problem of Cooper pairing, we take the polariton density $n_\mathrm{p}=3.103\times10^{11}\,\mbox{cm}^{-2}$ in the vicinity of the roton instability that occurs at the critical density $n_\mathrm{p}^\mathrm{crit}=3.1051\times10^{11}\,\mbox{cm}^{-2}$. However, for exciton-polaritons in QW, where the typical densities are or the order of $10^{10}\,\mbox{cm}^{-2}$, this value may be beyond the reach of current experiments.

For both material systems, the exciton-exciton interaction constant is taken $U=0.1\,\mu\mbox{eV}~\mu\mbox{m}^2$. The bare exciton-exciton interaction is usually assumed to be stronger, especially in QW (see, e.g.,~\cite{Rapaport}). However as was shown in~\cite{Maslova,Yamamoto}, the exciton-exciton interaction matrix element for indirect excitons is a decaying function of the transfer momentum, hence for momenta that lie in the region of the roton minimum that are of interest in the current work, the effective value of $U$ can be lower.

\section{Spectrum of excitations and Eliashberg functions}\label{Sec3}

\subsection{Roton minimum and instability}\label{sec_roton}

The denominator of Eq.~(\ref{V_ee}) determines the dispersions of excitations in the system, i.e. the real frequencies $\omega$ where the denominator vanishes and the interaction has poles at $i\omega_n\rightarrow\omega+i\delta$ (with $\delta\rightarrow+0$).

\begin{figure}[b]
\begin{center}
\includegraphics[width=\columnwidth]{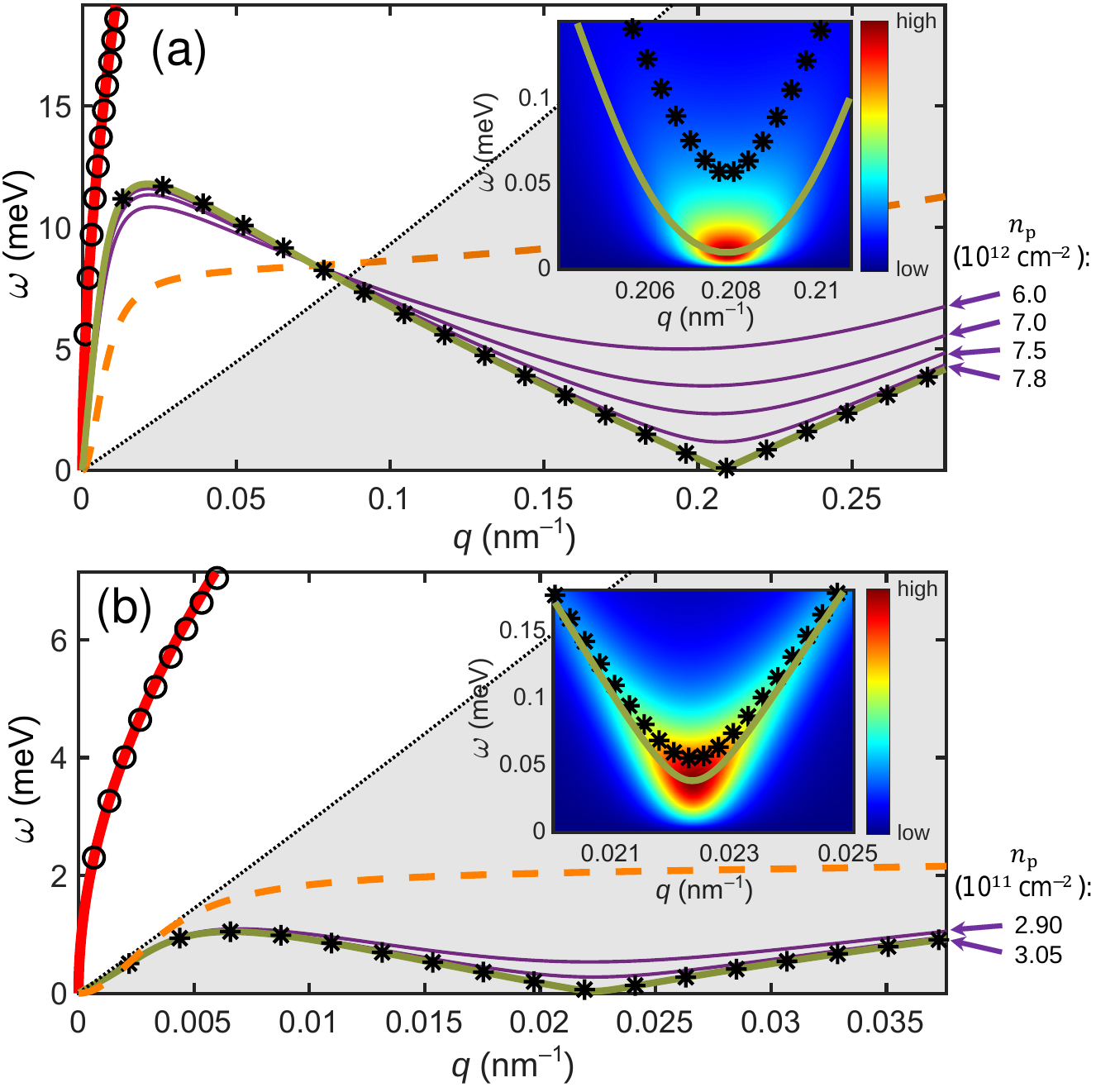}
\end{center}
\caption{\label{Fig_disp}Dispersions of excitations in the coupled electron-polariton system based on (a) TMD, at the polariton density $n_\mathrm{p}=7.8987\times10^{12}\,\mbox{cm}^{-2}$, and (b) QW, at $n_\mathrm{p}=3.103\times10^{11}\,\mbox{cm}^{-2}$. Circles and dashed line: excitations in individual layers, i.e. 2DEG plasmon and lower polariton. Red and olive solid lines: upper and lower bogolon-plasmon hybrid modes calculated in RPA. Stars: Bogoliubov mode dispersion calculated in TFA; thin purple lines show the same dispersions at progressively lower polariton densities $n_\mathrm{p}$ indicated on the right. The lower bogolon-plasmon hybrid modes, entering the single-particle continua (grey shaded areas), acquire finite lifetime. Insets show the magnified view of their dispersions in RPA and TFA near the roton minimum superimposed on the spectral density $-(1/\pi)\mathrm{Im}\,V_\mathrm{ee}(q,\omega+i\delta)$.}
\end{figure}

When the density responses are considered in the static limit of TFA ($q\to 0$, $\omega=0$), $\Pi_\mathrm{e}$
does not depend on frequency and one obtains a single pole $\omega=E_q^\mathrm{p}$
corresponding to the Bogoliubov dispersion of excitations [Eq.~(\ref{Bog_disp}) with the statically screened $\tilde{v}_q^\mathrm{xx}$].  As pointed out in Refs.~\cite{Shelykh,supersolid,Cotlet} addressing this limit, upon the increase of the polariton density $n_\mathrm{p}$ (or decrease of the interlayer distance $L$), the negative contribution to the renormalized exciton-exciton interaction $\tilde{v}_q^\mathrm{xx}$ (\ref{tilde_v_xx}) due to the 2DEG response starts to dominate over the positive terms, strongly softening the bogolon dispersion. Such softened dispersions are shown in Fig.~\ref{Fig_disp} by stars, as calculated for both TMD- [Fig.~\ref{Fig_disp}(a)] and QW-based [Fig.~\ref{Fig_disp}(b)] realisations at the high polariton densities $n_\mathrm{p}$ approaching the critical values $n_\mathrm{p}^\mathrm{crit}$. Taking progressively lower $n_\mathrm{p}$, as shown by purple lines, we can observe gradual disappearance of the roton minima. Thus the effect of interactions in the strongly-coupled Bose-Fermi system on the bare polariton dispersion (orange dashed lines in Fig.~\ref{Fig_disp}) is twofold: first, quadratic dispersion in the region of small momenta $q\rightarrow0$ becomes linear (``sound'' waves in a BEC), and, second, the dispersion becomes softened at higher momenta $q\sim1/2L$ (see Appendix~\ref{Appendix_A} for details).

When we take into account dynamical screening effects, $\Pi_\mathrm{e}$ becomes frequency-dependent, hence $\tilde{v}_q^\mathrm{xx}$ and the Bogoliubov spectrum (\ref{Bog_disp}) become frequency-dependent, too. In this case, one obtains two solutions
$i\omega_n=\omega_q^i$ of the dispersion equation. They correspond to the undamped upper ($i={\rm up}$) and lower ($i={\rm low}$) hybrid mode dispersions shown in Fig.~\ref{Fig_disp} by olive lines at the highest polariton densities $n_\mathrm{p}$. The lower hybrid mode becomes damped upon submerging into the single-particle continuum of electron-hole excitations in the 2DEG $\omega\leq(q^2+2qk_\mathrm{F})/2m$, which is marked by the shaded area in Fig.~\ref{Fig_disp}(a,b). In this regime the energy $\omega_q^\mathrm{low}$ of the damped mode is obtained by fitting the spectral density $-(1/\pi)\mathrm{Im}\,V_\mathrm{ee}(q,\omega+i\delta)$ with the Fano resonance line shape (see details discussed in a separate work \cite{inprep_letter}). In RPA we observe the same softening of the lower hybrid mode dispersion $\omega_q^\mathrm{low}$ due to the 2DEG-induced renormalization of exciton-exciton interaction as in TFA. The broadening of the lower hybrid mode inside the continuum is demonstrated in the insets of Fig.~\ref{Fig_disp} by overlaying the close-up view of its dispersion in the vicinity of the roton minimum on the spectral density $-(1/\pi)\mathrm{Im}\,V_\mathrm{ee}(q,\omega+i\delta)$. The dispersion of undamped bogolon calculated in TFA \cite{Cotlet} plotted in the same panels with star symbols demonstrates an upward energy shift. Note that in the QW-based case shown in Fig.~\ref{Fig_disp}(b), the lower hybrid mode is damped at all momenta.

\begin{figure}[t]
\begin{center}
\includegraphics[width=\columnwidth]{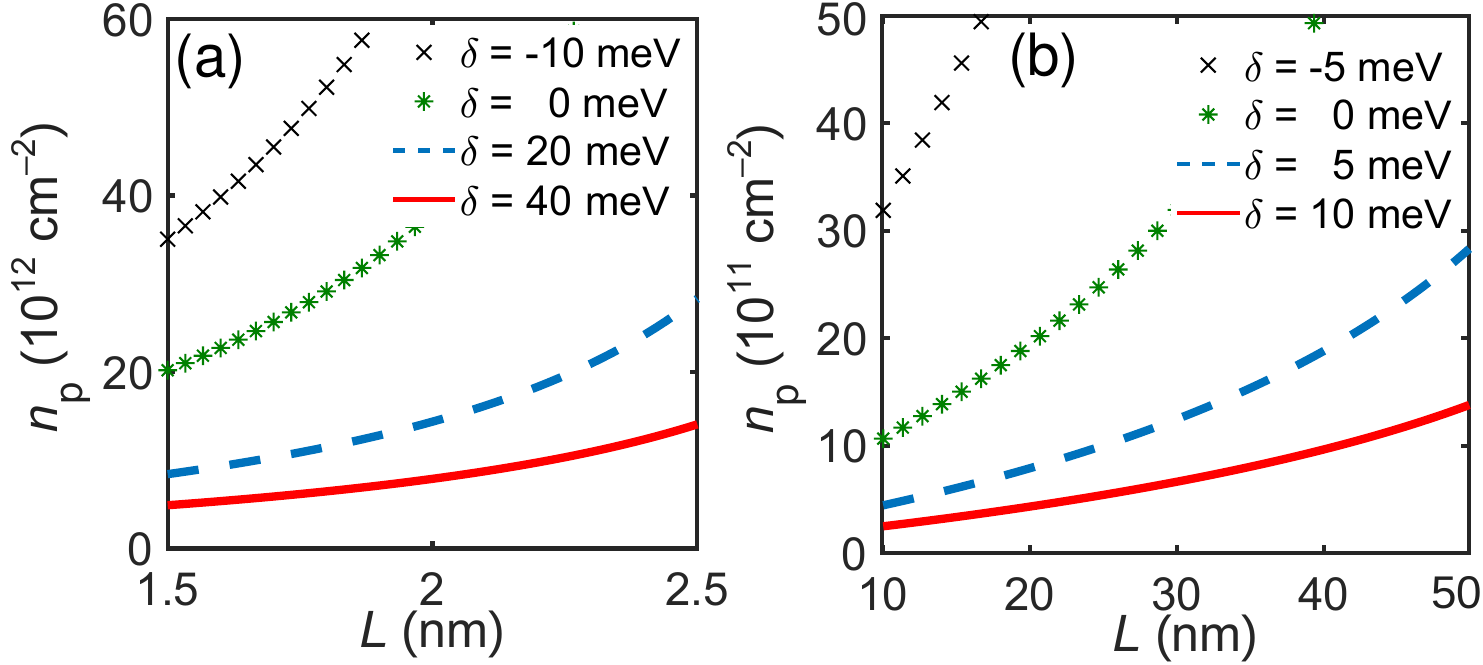}
\end{center}
\caption{\label{Fig_phase}Phase diagram of stability of the electron-polariton system based on TMD (a) and QW (b) in terms of the interlayer distance $L$ and polariton density $n_\mathrm{p}$ at different photon-to-exciton detunings $\delta$ shown in the legends. The system stays in a spatially uniform BEC state below the critical line and undergoes the roton instability above it.}
\end{figure}

Formation of the roton minimum is an important phenomenon determining the system stability \cite{Shelykh} and significantly contributing to the electron pairing interaction \cite{Cotlet}. Similar roton enhancement of the fermion pairing was predicted for atomic Bose-Fermi mixtures \cite{Dutta}. The system is stable when the roton minimum energy is positive, and the instability occurs whenever $E_q^\mathrm{p}(i\omega_n=0)=0$ at finite $q$. In Fig.~\ref{Fig_phase} we show the phase diagrams with the regions of a stable uniform BEC of polaritons (below the phase boundary line) and some other symmetry-breaking phase (above the line) resulting from the roton instability, which can be a supersolid state or a density wave \cite{Bland,Hertkorn,supersolid}. In this figure, in addition to the aforementioned values of detuning $\delta$ used in our superconductivity calculations, we chose several different values to demonstrate that the instability is achievable at lower $n_\mathrm{p}$ as $\delta$ is increased, since the lower polaritons become more excitonic and hence stronger interacting with the electrons. Our calculations (see below) show that the Cooper pairing in the 2DEG occurs only in a close vicinity of the unstable region due to the dominating contribution of low-energy roton excitations to electron-electron attraction.

\subsection{Eliashberg functions and coupling constants}\label{Sec_Eliashberg_func}

To demonstrate the role of virtual excitations in the effective electron-electron attraction resolved over the frequency $\omega$, we calculate the Eliashberg function
\begin{align}
\alpha^2F(\omega)&=-\frac{\mathcal{N}}\pi\langle\mathrm{Im}\,V_\mathrm{ee}(|\mathbf{k}-\mathbf{k}'|,\omega+i\delta)\rangle_\mathrm{FS}\nonumber\\
&=\alpha^2F_\mathrm{up}(\omega)+\alpha^2F_\mathrm{low}(\omega)+\alpha^2F_\mathrm{sp}(\omega).\label{alpha2F_ee}
\end{align}
Here $\mathcal{N}=m_\mathrm{e}^*/2\pi$ is the density of states at the Fermi level of the 2DEG, and $\langle\ldots\rangle_\mathrm{FS}$ denotes angular averaging with both momenta $\mathbf{k}$ and $\mathbf{k}'$ encircling the Fermi surface $|\mathbf{k}|=k_\mathrm{F}$ (here $k_\mathrm{F}=\sqrt{4\pi n_\mathrm{e}/g}$ is the Fermi momentum), so that $q=|\mathbf{k}-\mathbf{k}'|$ changes from 0 to $2k_\mathrm{F}$.

Using the spectral representation of the electron-electron interaction (\ref{V_ee})
\begin{multline}
V_\mathrm{ee}(q,i\omega_n) = v_q^\mathrm{ee}+\frac{2W_q^\mathrm{up}\omega_q^\mathrm{up}}{(i\omega_n)^2
-(\omega_q^\mathrm{up})^2}+\frac{2W_q^\mathrm{low}\omega_q^\mathrm{low}}{(i\omega_n)^2 -(\omega_q^\mathrm{low})^2}\\
-\frac1\pi\int_0^\infty\mathrm{Im}\,V_{\rm ee}(q,\nu)\frac{2\nu\:d\nu}{(i\omega_n)^2-\nu^2},\label{V_c_spectr2}
\end{multline}
in the second line of Eq.~(\ref{alpha2F_ee}) we have separated the contributions of the upper and lower hybrid modes, as well as that of the single-particle continuum. The spectral weights $W_q^\mathrm{up,low}$ are found from residues of $V_\mathrm{ee}$ at the corresponding poles $i\omega_n=\omega_q^\mathrm{up,low}$. When the lower hybrid mode enters the continuum, the third term in the right-hand side of (\ref{V_c_spectr2}) is absent so $W_q^\mathrm{low}$ is undefined, but $W_q^\mathrm{up}$ can always be found.

In parallel, we consider a similar spectral representation of the screened Coulomb interaction in the 2DEG (\ref{V_1}):
\begin{align}
V_1(q,i\omega_n)&=v_q^\mathrm{ee}+\frac{2W^\mathrm{pl}_q\omega^\mathrm{pl}_q}{(i\omega_n)^2-(\omega^\mathrm{pl}_q)^2}\nonumber\\
&-\frac1\pi\int_0^\infty\mathrm{Im}\,V_1(q,\nu)\frac{2\nu\:d\nu}{(i\omega_n)^2-\nu^2}.\label{V_1_spectr}
\end{align}
The terms in the right-hand side correspond to the unscreened Coulomb repulsion, attraction due to the 2DEG plasmons, and single-particle excitations. The dispersion $\omega^\mathrm{pl}_q$ of the 2DEG plasmons is found as poles of $V_1(q,i\omega)$, where $1-v^\mathrm{ee}_q\Pi_\mathrm{e}(q,\omega^\mathrm{pl}_q)=0$, and the spectral weight
$W^\mathrm{pl}_q$ is found from the residues of $V_1$ at these poles. Correspondingly, the Eliashberg function for an isolated 2DEG,
\begin{align}
\alpha^2F_1(\omega)&=-\frac{\mathcal{N}}\pi\langle\mathrm{Im}\,V_1(|\mathbf{k}-\mathbf{k}'|,\omega+i\delta)\rangle_\mathrm{FS}\nonumber\\
&=\alpha^2F_\mathrm{pl}(\omega)+\alpha^2F_\mathrm{sp}(\omega),\label{alpha2F_1}
\end{align}
is separated into contributions of plasmons and single-particle excitations. Note that the difference between the Eliashberg functions (\ref{alpha2F_ee}) and (\ref{alpha2F_1}) is caused by the interaction $V_2=V_\mathrm{ee}-V_1$ (\ref{V_2}) of the 2DEG with the polariton subsystem, so we assume that the contribution of the single-particle continuum $\alpha^2F_\mathrm{sp}(\omega)$ to both $\alpha^2F(\omega)$ and $\alpha^2F_1(\omega)$ is the same. The contribution of the lower hybrid mode $\alpha^2F_\mathrm{low}(\omega)$ to (\ref{alpha2F_ee}) is attributed in our approach to both undamped and damped regions of its dispersion.

\begin{figure}[t]
\begin{center}
\includegraphics[width=0.7\columnwidth]{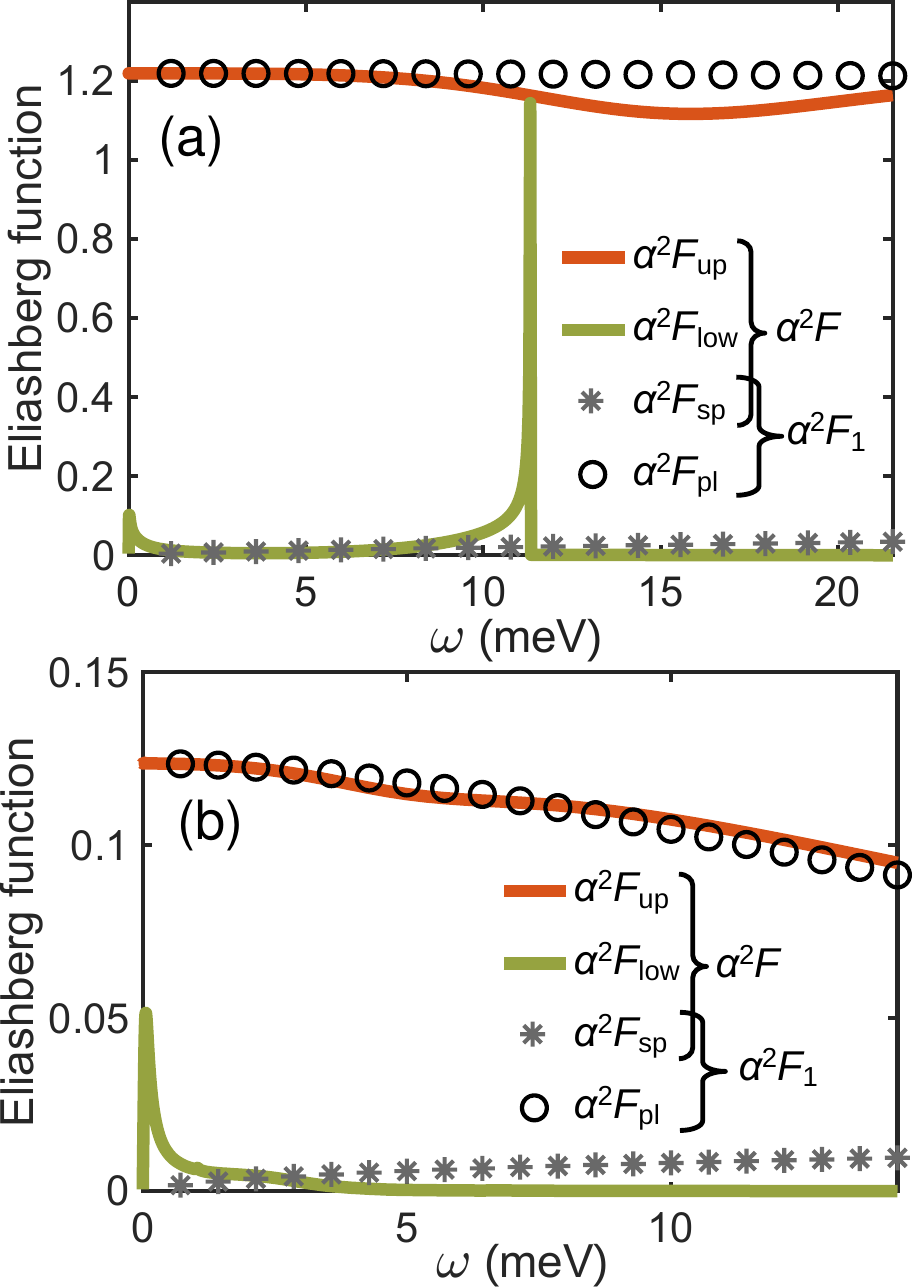}
\end{center}
\caption{\label{Fig_Eliashberg}Eliashberg functions of the screened electron-electron interaction in the 2DEG for (a) TMD- and (b) QW-based systems, separated into different parts. The Eliashberg function $\alpha^2F$ (\ref{alpha2F_ee}) in the presence of the polariton layer is divided into contributions of the upper hybrid mode ($\alpha^2F_\mathrm{up}$, red solid line), lower hybrid mode both outside and inside the continuum ($\alpha^2F_\mathrm{low}$, olive solid line), and single-particle continuum of the 2DEG ($\alpha^2F_\mathrm{sp}$, stars). The Eliashberg function for 2DEG in the absence of polariton layer $\alpha^2F_1$ (\ref{alpha2F_1}) consists of the contributions of the same single particle continuum $\alpha^2F_\mathrm{sp}$ and of the plasmons ($\alpha^2F_\mathrm{1,pl}$, circles). The contributions of the depicted functions to the coupling constants are shown in Table~\ref{Table}.}
\end{figure}

Fig.~\ref{Fig_Eliashberg} shows the parts of the Eliashberg functions  (\ref{alpha2F_ee}) and (\ref{alpha2F_1})
calculated in RPA under the conditions of Fig.~\ref{Fig_disp}. The dominating contributions are provided, respectively, by the plasmon and by the upper hybrid mode which corresponds to a slightly modified plasmon, too. The attractive interactions due to these modes severely compete with the bare Coulomb repulsion in the electron gas \cite{Rietschel}. Contribution of the single-particle continuum is smooth and small in magnitude, while that of the lower hybrid mode has peaks corresponding to extrema of its dispersion: the maximum at $\omega\sim\Omega_\mathrm{R}$ (it is almost unseen in the QW case since it is reached inside the continuum and thus subject to broadening) and the roton minimum at very low energies.

The Eliashberg function determines the dimensionless coupling constant
\begin{equation}
\lambda=2\int_0^\infty\frac{\alpha^2F(\omega)}\omega \:d\omega,\label{lambda}
\end{equation}
which is relevant to superconductivity and Fermi-liquid properties. Separating $\alpha^2F(\omega)$ into physically distinct parts (\ref{alpha2F_ee}) allows to single out corresponding contributions to $\lambda$ listed in the upper part of Table~\ref{Table}. The lower hybrid mode provides rather large contribution since the integral (\ref{lambda}) is dominated by the vicinity of the low-energy roton minimum due to the factor $1/\omega$. The upper hybrid mode, corresponding to the renormalized 2DEG plasmon with the dispersion $\omega_q^\mathrm{up}\propto\sqrt{q}$ and the spectral weight $W_q^\mathrm{up}\propto1/\sqrt{q}$ at $q\rightarrow0$ provides $\alpha^2F_\mathrm{up}(\omega)\neq0$ in the limit $\omega\rightarrow0$, which makes its contribution to $\lambda$ formally divergent. Physically, this divergent effect of attraction at the Fermi surface due to renormalized plasmons should be compensated by the opposite divergence of Coulomb repulsion (see also Appendix~\ref{Appendix_B}). However, the Eliashberg function is not intended to properly take into account such compensation that requires stepping away from the Fermi surface in momentum space.

\begin{table}[t]
\begin{tabular}{cll}
\hline\hline
Contribution to $\lambda$ from&\;TMD&\;\;QW\\
\hline
$\alpha^2F_\mathrm{up}$&$\hphantom{-}\infty$&$\hphantom{-}\infty$\\
$\alpha^2F_\mathrm{sp}$&$\hphantom{-}0.15$&$\hphantom{-}0.06$\\
$\alpha^2F_\mathrm{low}$&$\hphantom{-}0.45$&$\hphantom{-}0.20$\\
\hline
\hline
Contribution to $\lambda_2$ from&\;TMD&\;\;QW \\
\hline
$\alpha^2F_\mathrm{up}-\alpha^2F_\mathrm{pl}$&$-0.038$&$-0.001$\\
$\alpha^2F_\mathrm{low}$&$\hphantom{-}0.45$&$\hphantom{-}0.20$\\
\hline\hline
\end{tabular}
\caption{Contributions to the coupling constants $\lambda$ (\ref{lambda}) and $\lambda_2$ (\ref{lambda_2}) from different terms in the Eliashberg functions $\alpha^2F$ (\ref{alpha2F_ee}) and $\alpha^2F_2$ (\ref{alpha2F_2}), calculated for the TMD- and QW-based systems under the conditions of Fig.~\ref{Fig_disp}.}
\label{Table}
\end{table}

To avoid the clash of divergences on the Fermi surface, we focus only on part $V_2$ of the total electron-electron interaction $V_\mathrm{ee}=V_1+V_2$ (\ref{V_ee_decomp}), which is caused by the presence of the polariton subsystem. In the absence of polaritons, the screened electron-electron repulsion $V_1$ (\ref{V_1}) in the 2DEG can be described by the effective Coulomb pseudopotential $\mu$. This quantity has a meaning of the dimensionless product of $\mathcal{N}$ and $V_1$, the latter averaged over the Fermi surface. It is the change $V_2=V_\mathrm{ee}-V_1$ (\ref{V_2}) of the interaction due to polaritons which is physically the most meaningful for the problem of polariton-mediated superconductivity, hence we consider its Eliashberg function [i.e. difference of (\ref{alpha2F_ee}) and (\ref{alpha2F_1})] separately:
\begin{align}
\alpha^2F_2(\omega)&=-\frac{\mathcal{N}}\pi\langle\mathrm{Im}\,V_2(|\mathbf{k}-\mathbf{k}'|,\omega+i\delta)\rangle_\mathrm{FS}\nonumber\\
&=\alpha^2F_\mathrm{up}(\omega)-\alpha^2F_\mathrm{pl}(\omega)+\alpha^2F_\mathrm{low}(\omega).\label{alpha2F_2}
\end{align}
This function refers to the pairing interaction induced in the 2DEG by the polariton layer. The corresponding coupling constant
\begin{equation}
\lambda_2=2\int_0^\infty\frac{\alpha^2F_2(\omega)}\omega\:d\omega,\label{lambda_2}
\end{equation}
similarly to (\ref{lambda}), consists of physically distinct contributions, shown in the lower part of Table~\ref{Table}. The first two terms in the right-hand side of (\ref{alpha2F_2}) provide together a finite (and very small) contribution to $\lambda_2$ because in the limit $q\rightarrow0$ both the 2DEG plasmon and the upper hybrid mode have the same dispersions $\omega_q^\mathrm{pl}\approx\omega_q^\mathrm{up}\propto\sqrt{q}$ and spectral weights $W_q^\mathrm{pl}\approx W_q^\mathrm{up}\propto1/\sqrt{q}$, leading to $\alpha^2F_\mathrm{pl}(\omega)\approx\alpha^2F_\mathrm{up}(\omega)$ at $\omega\rightarrow0$ (compare the circles and the red lines in Fig.~\ref{Fig_Eliashberg}). The overwhelming contribution to both $\lambda$ and $\lambda_2$ is provided by the lower hybrid mode $\alpha^2F_\mathrm{low}(\omega)$ due to its softness near the roton minimum (olive lines in Fig.~\ref{Fig_Eliashberg}).

\section{Theory of superconductivity}\label{Sec4}

\subsection{Momentum and frequency resolved gap equations}

We analyze the electron Cooper pairing in the 2DEG starting from the Gor'kov equations given in the diagrammatic form in Fig.~\ref{Fig_Diag}(f):
\begin{equation}
\hat{G}=\hat{G}_0+\hat{G}_0\hat\Sigma\hat{G}.\label{Gorkov_eq}
\end{equation}
Here the Nambu notation is used for the matrix Green's function of electrons $\hat{G}(\mathbf{k},\tau)=-\langle T_\tau \hat{c}_\mathbf{k}(\tau)\hat{c}^+_\mathbf{k}(0)\rangle$ defined in terms of Nambu spinors  $\hat{c}_\mathbf{k}=(c_{\mathbf{k}\uparrow},c^+_{-\mathbf{k}\downarrow})^T$. The self-energy
\begin{equation}
\hat\Sigma(\mathbf{k},i\varepsilon_n) \!=\! -T\!\sum_{\mathbf{k}'\varepsilon_m}\!V_\mathrm{ee}(|\mathbf{k}\!- \!\mathbf{k}'|,i\varepsilon_n\!-\!i\varepsilon_m)\sigma_z\hat{G}(\mathbf{k}'\!,i\varepsilon_m)\sigma_z\label{Sigma}
\end{equation}
is found self-consistently from $\hat{G}$; here $i\varepsilon_n=\pi i(n+1/2)T$ are the fermionic Matsubara frequencies. Decomposing the self-energy in a usual way over the Pauli matrices $\hat\Sigma=i\varepsilon_n(1-Z)+\sigma_x\varphi+\sigma_z\chi$ (the $\sigma_y$ term can be eliminated by a gauge transformation) and taking into account that the non-interacting electron Green's function is given by  $\hat{G}_0(\mathbf{k},i\varepsilon_n)=(i\varepsilon_n+\mu_\mathrm{e}-\sigma_z\epsilon_k^\mathrm{e})^{-1}$, we find the solution of the Gor'kov equations
\begin{equation}
\hat{G}(\mathbf{k},i\varepsilon_n)=\frac{i\varepsilon_nZ+\sigma_x\varphi+\sigma_z\xi}{(i\varepsilon_nZ)^2-\varphi^2-\xi^2},\label{G}
\end{equation}
where $\xi=\epsilon_k^\mathrm{e}-\mu_\mathrm{e}+\chi$ is the renormalized electron energy counted from the chemical potential $\mu_\mathrm{e}\approx E_\mathrm{F}=k_\mathrm{F}^2/2m_\mathrm{e}^*$ of the 2DEG. The functions $Z$, $\varphi$, $\chi$ depend on $(\mathbf{k},i\varepsilon_n)$ and are responsible for renormalization of quasiparticles in a superconducting state, appearance of an energy gap, and the interaction-induced change of the electron dispersion, respectively.

Substitution of Eq.~(\ref{G}) into Eq.~(\ref{Sigma}) yields the self-consistent gap equation. As adopted in the Eliashberg theory~\cite{Marsiglio,Eliashberg,Scalapino,McMillan}, we assume that the renormalization $\chi$ of the quasiparticle dispersion near the Fermi surface ($|\mathbf{k}|=k_\mathrm{F}$, $\varepsilon_n=0$) is smooth and hence can be absorbed into $\epsilon_k^\mathrm{e}-\mu_\mathrm{e}$, so that $\chi$ can be omitted. Linearizing the gap equation at $T=T_\mathrm{c}$ when $\varphi\rightarrow0$, and projecting it on the unity matrix and on $\sigma_x$, we obtain the set of equations:
\begin{align}
\varepsilon_n[1-Z(\mathbf{k},i\varepsilon_n)]&=-\frac{T}S\sum_{\mathbf{k}'\varepsilon_m}V_\mathrm{ee}(|\mathbf{k}-\mathbf{k}'|,i\varepsilon_n-i\varepsilon_m)\nonumber\\
&\times\frac{\varepsilon_mZ(\mathbf{k}',i\varepsilon_m)}{[i\varepsilon_mZ(\mathbf{k}',i\varepsilon_m)]^2-\xi_{k'}^2}, \quad
\label{Z_eq}
\\
\varphi(\mathbf{k},i\varepsilon_n)&=\frac{T}S\sum_{\mathbf{k}'\varepsilon_m}V_\mathrm{ee}(|\mathbf{k}-\mathbf{k}'|,i\varepsilon_n-i\varepsilon_m)\qquad\qquad
\nonumber\\
&\times\frac{\varphi(\mathbf{k}',i\varepsilon_m)}{[i\varepsilon_mZ(\mathbf{k}',i\varepsilon_m)]^2-\xi_{k'}^2},\quad
\label{phi_eq}
\end{align}
where $\xi_{k'}=\epsilon^\mathrm{e}_{k'}-\mu_\mathrm{e}$.

\begin{figure*}
\begin{center}
\includegraphics[width=0.9\textwidth]{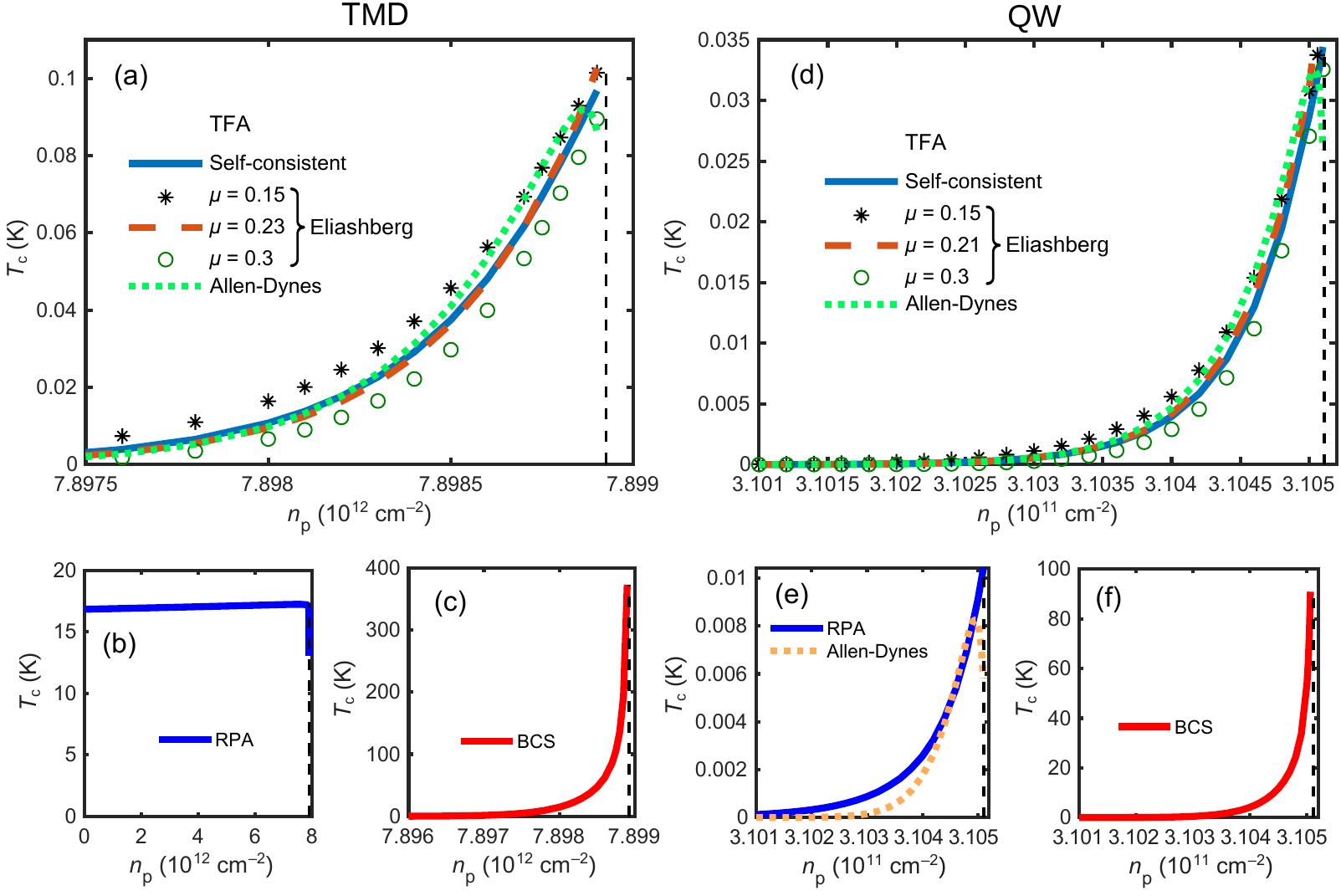}
\end{center}
\caption{\label{Fig_Tc}Superconducting critical temperatures $T_\mathrm{c}$ in TMD- [panels (a--c) on the left-hand side] and QW-based [panels (d--f) on the right-hand side] systems as functions of the polariton density $n_\mathrm{p}$ in the vicinity of the roton instability (vertical dashed lines). (a,d) Calculations with the screening in TFA. Solutions of the self-consistency Eqs.~(\ref{Z_eq})--(\ref{phi_eq}) with the full frequency-momentum dependence are shown by the blue solid lines; the dashed red lines, stars, and circles show solutions of the Eliashberg Eqs.~(\ref{Z_Eliashberg_reg})--(\ref{phi_Eliashberg_reg}) at different Coulomb pseudopotentials $\mu$ shown in the legends. Results of the Allen-Dynes formula are shown by the dotted lines. (b,e) Calculations with the screening in RPA. Solutions of the self-consistency Eqs.~(\ref{Z_eq})--(\ref{phi_eq}) with full frequency-momentum dependence are shown by the blue solid lines, and the result of the Allen-Dynes formula for the QW-based system is shown by the orange dotted line.
(c,f) Results of the BCS approach (\ref{BCS}) with the statically screened interaction.}
\end{figure*}

The problem of superconductivity in the 2DEG admits several different approaches that we employ and compare below. The first one is the direct numerical solution of Eqs.~(\ref{Z_eq})--(\ref{phi_eq}) taking into account both the momentum and frequency dependencies of $Z$ and $\varphi$. The first equation (\ref{Z_eq}) is solved iteratively to obtain $Z$ which is then substituted into the second equation (\ref{phi_eq}). The latter is solved as the eigenfunction problem for $\varphi$ with the eigenvalue 1. The ranges of both $k$ and $\varepsilon_n$ are separated into sufficiently small (until the results converge) intervals in which the functions $Z$, $\varphi$, $V_\mathrm{ee}$ can be treated as constants, and the high-frequency tails of the sums over $\varepsilon_n$ are approximated with the integrals. In Appendix \ref{Appendix_B} we provide more details about the frequency and momentum dependence of $Z$  and $\varphi$.

The results of this direct numerical solution depend on the approximation describing the interaction screening. In Fig.~\ref{Fig_Tc}(a,d), the solid lines show the results of this approach when $V_\mathrm{ee}$ is screened in TFA. Any noticeable pairing exists only in a very close vicinity of the roton instability, where the polariton density $n_\mathrm{p}$ is no more than 0.01\% smaller than the critical value (indicated by the vertical dashed lines). This circumstance was clearly seen during analysis of Fig.~\ref{Fig_Eliashberg} and Table~\ref{Table}, where the overwhelming contribution to the low-frequency Eliashberg function and hence to the coupling constant (\ref{lambda}) was provided by the hybrid Bogoliubov excitations near the roton minimum. Similar conclusions were made in Ref.~\cite{Cotlet}, where only TFA was used to account for the interaction screening.

Considering the interaction $V_\mathrm{ee}$ in RPA, which takes into account the dynamical effects such as plasmon-bogolon hybridization and damping of the lower hybrid mode near the roton minimum, yields controversial results displayed
in Fig.~\ref{Fig_Tc}(b,e). The critical temperature for the TMD-based system $T_\mathrm{c}\approx17\,\mbox{K}$ [see Fig.~\ref{Fig_Tc}(b)] turns out to be rather high. Furthermore, the Cooper pairing exists in this approach even in the absence of polaritons (at $n_\mathrm{p}=0$), i.e. in an isolated 2DEG, merely due to plasmon-mediated electron attraction. Similar artefact of the RPA-screened Coulomb interaction was discovered in a three-dimensional electron gas (3DEG) in the strongly-interacting regime \cite{Rietschel}. Inclusion of vertex corrections (or local field factors) which provide spin-selective weakening of the screening at small distances generally diminishes or excludes the plasmon-mediated pairing \cite{Takada,Richardson} in the 3DEG. Thus we conclude that in the strongly-interacting regime, which emerges in the considered TMD-based system, RPA produces unreliable results. Indeed, the relative strength $r_\mathrm{s}=m_\mathrm{e}^*e^2/\varepsilon\hbar^2\sqrt{\pi n_\mathrm{e}}$ of Coulomb interaction in the 2DEG is markedly larger than unity, $r_\mathrm{s}=3.8$, at TMD-based parameters. In contrast, for the QW-based system [Fig.~\ref{Fig_Tc}(e)], $T_\mathrm{c}$ obtained in RPA is about 3.5 times lower than that in TFA [Fig.~\ref{Fig_Tc}(d)], and the pairing exists again only in a close vicinity of the critical polariton density. The spurious plasmon-mediated pairing does not develop in this case because of the generally weaker electron-electron and electron-exciton interaction, $r_\mathrm{s}=0.55$.

\subsection{Eliashberg equations}\label{Sec_Eliashberg}

The second approach to solve the self-consistency equations (\ref{Z_eq})--(\ref{phi_eq}) is the Eliashberg theory \cite{Eliashberg,Marsiglio,McMillan,Scalapino}, where the momentum dependence of $Z$ and $\varphi$ is assumed to be smooth in the vicinity of the Fermi surface, so that $Z(\mathbf{k},i\varepsilon_n)\approx Z(k_\mathrm{F},i\varepsilon_n)\equiv Z(i\varepsilon_n)$, and the same for $\varphi(i\varepsilon_n)$. This assumption allows us to perform summation over $\mathbf{k}$ in Eqs.~(\ref{Z_eq})--(\ref{phi_eq}) according to the rule
\begin{multline}
\frac1S\!\sum_{\mathbf{k}'}\!V_\mathrm{ee}(|\mathbf{k}-\mathbf{k}'|,i\varepsilon_n-i\varepsilon_m)\frac{f(k')}{[i\varepsilon_mZ(\mathbf{k}',i\varepsilon_m)]^2-\xi_{k'}^2}\\
\approx-\pi\mathcal{N}V_\mathrm{ee}^\mathrm{FS}(i\varepsilon_n,i\varepsilon_m)\frac{f(k_\mathrm{F})}{|\varepsilon_m|Z(i\varepsilon_m)},
\end{multline}
where
$V_\mathrm{ee}^\mathrm{FS}(i\varepsilon_n,i\varepsilon_m)=\langle V_\mathrm{ee}(|\mathbf{k}-\mathbf{k}'|,i\varepsilon_n-i\varepsilon_m)\rangle_\mathrm{FS}$. In the resulting set of Eliashberg equations
\begin{align}
\varepsilon_n[1-Z(i\varepsilon_n)] \!=\! \pi\mathcal{N}T\sum_{\varepsilon_m}V_\mathrm{ee}^\mathrm{FS}(i\varepsilon_n,i\varepsilon_m)\,\mathrm{sgn}(\varepsilon_m),
\label{Z_Eliashberg}
\\
\varphi(i\varepsilon_n) \!=\! -\pi\mathcal{N}T\sum_{\varepsilon_m}V_\mathrm{ee}^\mathrm{FS}(i\varepsilon_n,i\varepsilon_m)\frac{\varphi(i\varepsilon_m)}{|\varepsilon_m|Z(i\varepsilon_m)},
\label{phi_Eliashberg}
\end{align}
the first Eq.~(\ref{Z_Eliashberg}) is actually
an explicit expression for $Z(i\varepsilon_n)$, while the second Eq.~(\ref{phi_Eliashberg}) should be regularized in order to take into account the high-frequency tail of $V_\mathrm{ee}$ which provides the divergent contribution to the frequency sum. Conventional approach \cite{Scalapino,McMillan,Marsiglio} here is to separate the dimensionless interaction $\mathcal{N}V_\mathrm{ee}^\mathrm{FS}$ into the repulsive Coulomb pseudopotential $\mu$ which is almost constant in a broad frequency range of the order of the characteristic 2DEG energy $E_\mathrm{F}$, and the attractive pairing contribution $K(i\varepsilon_n-i\varepsilon_m)$ which is essentially dynamical and exists below the ``Debye frequency'' $\omega_\mathrm{D}$, i.e. $\mathcal{N}V_\mathrm{ee}^\mathrm{FS}(i\varepsilon_n,i\varepsilon_m)=\mu\Theta(E_\mathrm{F}-|\varepsilon_n|)\Theta(E_\mathrm{F}-|\varepsilon_m|)-K(i\varepsilon_n-i\varepsilon_m)$. Here $\omega_\mathrm{D}$ can be understood as the highest energy of virtual excitations in the polariton system, which is close to $\Omega_\mathrm{R}$. Due to the factor $\mathrm{sgn}(\varepsilon_m)$, $\mu$ is canceled in (\ref{Z_Eliashberg}), while
(\ref{phi_Eliashberg}) takes the form
\begin{align}
\varphi(i\varepsilon_n)=\pi T\!\!\!\!\sum_{|\varepsilon_m|<\omega_\mathrm{D}}\!\!\!\!\{K(i\varepsilon_n-i\varepsilon_m)-\mu\}\frac{\varphi(i\varepsilon_m)}{|\varepsilon_m|Z(i\varepsilon_m)}\nonumber\\
-\mu\varphi_\infty\log(E_\mathrm{F}/\omega_\mathrm{D}).\label{phi_Eliashberg1}
\end{align}
The last term comes from the high-frequency sum
$\sum_{\omega_\mathrm{D}<|\varepsilon_m|<E_\mathrm{F}}\varphi(i\varepsilon_m)/|\varepsilon_m|Z(i\varepsilon_m)$ transformed into an integral with $\varphi(i\varepsilon_m)\approx\varphi_\infty$ and $Z(i\varepsilon_m)\approx1$ at $|\varepsilon_m|>\omega_D$.
Taking high frequencies $\omega_\mathrm{D}<|\varepsilon_n|<E_\mathrm{F}$ in Eq.~(\ref{phi_Eliashberg1}), we obtain $\varphi_\infty$ and substitute it back into Eq.~(\ref{phi_Eliashberg1}) taken at low frequencies $|\varepsilon_n|<\omega_\mathrm{D}$, which yields:
\begin{equation}
\varphi(i\varepsilon_n)=\pi T\!\!\!\!\sum_{|\varepsilon_m|<\omega_\mathrm{D}}\!\!\!\!\{K(i\varepsilon_n-i\varepsilon_m)-\mu^*\}\frac{\varphi(i\varepsilon_m)}{|\varepsilon_m|Z(i\varepsilon_m)},\label{phi_Eliashberg3}
\end{equation}
where
\begin{equation}
\mu^*=\frac\mu{1+\mu\log(E_\mathrm{F}/\omega_\mathrm{D})}\label{mu_ren}
\end{equation}
is the Coulomb pseudopotential renormalized when being transferred from the high-frequency $\omega_\mathrm{D}<|\varepsilon_m|<E_\mathrm{F}$ to the low-frequency $|\varepsilon_m|<\omega_\mathrm{D}$ region. The substitution $\mu^*\rightarrow\mu^*\Theta(\omega_\mathrm{D}-|\varepsilon_m|)$ in Eq.~(\ref{phi_Eliashberg3}) allows to safely extend the frequency summation back to infinity, so that the final set of regularized Eliashberg equations reads:
\begin{equation}
\varepsilon_n[1-Z(i\varepsilon_n)] = \pi T\sum_{\varepsilon_m}K(i\varepsilon_n-i\varepsilon_m)\,\mathrm{sgn}(\varepsilon_m),
\label{Z_Eliashberg_reg}
\end{equation}
\begin{multline}
\varphi(i\varepsilon_n) =\pi T\sum_{\varepsilon_m}\{K(i\varepsilon_n-i\varepsilon_m)-\mu^*\Theta(\omega_\mathrm{D}-|\varepsilon_m|)\}\\  \times\frac{\varphi(i\varepsilon_m)}{|\varepsilon_m|Z(i\varepsilon_m)}.
\label{phi_Eliashberg_reg}
\end{multline}

This set of equations is solved numerically by grouping the frequencies $\varepsilon_n$ to intervals where $\varphi(i\varepsilon_n)$ is approximately constant, and performing analytical integration of high-frequency tails of the frequency sums. The Coulomb pseudopotential $\mu$ can be considered as a single parameter characterizing the strength of Coulomb repulsion $V_1$, which exists in the 2DEG in the absence of polaritons as well. On the other hand, the pairing interaction $K(i\omega_n)$, which is responsible for the Fermi-surface-averaged polariton-mediated interaction $V_2$ (\ref{V_2}), relates to its Eliashberg function (\ref{alpha2F_2}) as
\begin{equation}
K(i\omega_n)\!=\!-\mathcal{N}\langle V_2(|\mathbf{k}-\mathbf{k}'|,i\omega_n)\rangle_\mathrm{FS}\!=\!\!\int\limits_0^\infty\!\!\alpha^2F_2(\nu)\frac{2\nu\:d\nu}{\omega_n^2\!+\!\nu^2}.
\end{equation}
To estimate $\omega_\mathrm{D}$ in Eq.~(\ref{mu_ren}), we use the McMillan's recipe \cite{McMillan,AllenDynes}
\begin{equation}
\omega_\mathrm{D}=\left.\int_0^\infty\alpha^2F_2(\nu)\:d\nu\right/\int_0^\infty\frac{\alpha^2F_2(\nu)}\nu\:d\nu,
\end{equation}
which provides $\omega_\mathrm{D}\approx0.025E_\mathrm{F}$ and $0.005E_\mathrm{F}$ for the TMD and QW cases, respectively.

The results for $T_\mathrm{c}$ obtained from the regularized Eliashberg equations (\ref{Z_Eliashberg_reg})--(\ref{phi_Eliashberg_reg}) are shown in Fig.~\ref{Fig_Tc}(a,d) for different Coulomb pseudopotentials $\mu$. Typical values of $\mu$ for conventional superconductors are in the range 0.1-0.4 \cite{Kawamura}, and our calculations using the expression $\mu=\mathcal{N}\langle V_1(|\mathbf{k}-\mathbf{k}'|,i\omega_n=0)\rangle_\mathrm{FS}$ with the screening taken in TFA provide $\mu=0.23$ for TMD and $\mu=0.21$ for QW [see dashed red lines in Fig.~\ref{Fig_Tc}(a,d)]. Results of the Eliashberg approach turn out to be surprisingly close to the solution of the self-consistency equations (\ref{Z_eq})--(\ref{phi_eq}) with the full frequency-momentum dependence, although its underlying presumption of smooth momentum dependence near the Fermi surface is not valid, as shown in Appendix~\ref{Appendix_B}. Taking slightly higher or lower $\mu$ results in respective decrease or increase of $T_\mathrm{c}$ [stars and circles in Fig.~\ref{Fig_Tc}(a,d)].

\subsection{Allen-Dynes formula}

Instead of solving the Eliashberg equations numerically, one can rely on the  Allen-Dynes formula \cite{AllenDynes}, the improved version of the McMillan formula \cite{McMillan} fitting the numerical solutions for a number of conventional superconductors. It provides $T_\mathrm{c}$ in terms of the coupling constant $\lambda$, renormalized Coulomb pseudopotential $\mu^*$, and several moments of the Eliashberg function $\alpha^2F(\omega)$. The latter can be taken in two approximations: (i) $\alpha^2F(\omega)$ (\ref{alpha2F_ee}) for the total electron-electron interaction $V_\mathrm{ee}$ screened in TFA, (ii) $\alpha^2F_2(\omega)$ (\ref{alpha2F_2}) for the polariton-induced contribution $V_2$ to the electron-electron interaction calculated in RPA. In both cases we obtain close values of the coupling constants $\lambda$ (\ref{lambda}) and $\lambda_2$ (\ref{lambda_2}). The critical temperatures $T_\mathrm{c}$ obtained in this calculation are plotted in Fig.~\ref{Fig_Tc}(a,d) [with the approximation (i)] and  in Fig.~\ref{Fig_Tc}(e) [with the approximation (ii), but only for the QW-based setup where RPA works adequately]. The Coulomb pseudopotentials $\mu$ were calculated, as in the previous section, by averaging the TFA-screened Coulomb interaction $V_1$ over the Fermi surface, and then renormalized to $\mu^*$ using Eq.~(\ref{mu_ren}).

In both cases, the Allen-Dynes formula provides $T_\mathrm{c}$ of the same order as the results of numerical solution of Eliashberg equations with the same approximation for the screening: TFA in Fig.~\ref{Fig_Tc}(a,d) and RPA in Fig.~\ref{Fig_Tc}(e). Although the coupling constants are similar in these two cases, $T_\mathrm{c}$ for QW in RPA [solid line in Fig.~\ref{Fig_Tc}(e)] is several times lower than in TFA [dotted line in Fig.~\ref{Fig_Tc}(d)] due to the smaller pre-exponential factor (or characteristic energy of the pairing excitations) $\sim\omega_\mathrm{D}$ which is determined by the roton minimum energy in RPA and by a higher energy $\Omega_\mathrm{R}$ in TFA.

\subsection{Bardeen-Cooper-Schrieffer approach}\label{Sec_BCS}

Finally, to solve the Cooper pairing problem one can also use the BCS approach, where the dynamical effects are neglected and only the momentum dependencies of the gap and interaction are accounted for. To apply this approach we take $Z=1$ in Eq.~(\ref{phi_eq}), replace $V_\mathrm{ee}$ by the statically screened interaction, $V_\mathrm{ee}(|\mathbf{k}-\mathbf{k}'|,i\varepsilon_n-i\varepsilon_m)\rightarrow V_\mathrm{ee}(|\mathbf{k}-\mathbf{k}'|,0)$, and assume $\varphi$ to be frequency-independent. After summation over $\varepsilon_n$, we obtain the linearized BCS gap equation
\begin{equation}
\varphi(\mathbf{k})=-\frac1S\sum_{\mathbf{k}'}V_\mathrm{ee}(|\mathbf{k}-\mathbf{k}'|,0)\frac{\varphi(\mathbf{k}')}{2|\xi_{k'}|}\tanh\frac{|\xi_{k'}|}{2T},\label{BCS}
\end{equation}
which is solved numerically as the eigenfunction problem. The resulting BCS values of $T_\mathrm{c}$, shown in Fig.~\ref{Fig_Tc}(c,f), are orders of magnitude higher than the critical temperatures obtained in the Eliashberg approach, and can even reach room temperature. We note that using the BCS approach, other authors obtained similar overestimated $T_\mathrm{c}$ \cite{Laussy_PRL,Laussy_JNP,Cherotchenko,Sedov,Sun_2DMater,Sun_Thesis,Sun_NJP}, therefore we conclude that neglection of the dynamical effects in the analysis of superconducting pairing is misleading.

In Appendix~\ref{Appendix_C}, we investigate the origin of large discrepancy between the predictions of the Eliashberg and BCS approaches described above. On a simple example of the TFA-screened interaction dominated by the roton exchange in vicinity of the roton instability, we demonstrate that the Eliashberg and BCS approaches implicitly assume different energy widths of the pairing region originating from the energy and momentum of rotons, respectively. Since the energy width involved  in the BCS calculations is 2-3 orders of magnitude larger than in the Eliashberg description, the predicted critical temperature turns out to be proportionally larger.

\section{Conclusions}\label{Sec5}

The hybrid Bose-Fermi system of exciton-polaritons in the Bose-condensed regime coupled to a 2DEG via an interlayer electron-exciton interaction was considered using the many-body quantum theory, with a special attention paid to the problem of potential superconductivity induced in the 2DEG in the presence of a polariton BEC \cite{Laussy_JNP,Cherotchenko,Sedov,Cotlet,Sun_2DMater,Sun_Thesis,Sun_NJP,Sun_PRR,Skopelitis}. We analyzed the electron-electron interaction undergoing combined screening by density responses of the polariton BEC and the 2DEG, the latter taken both in the Thomas-Fermi and random-phase approximations. With taking into account dynamical effects, hybridization of Bogoliubov quasiparticles of the polariton system with plasmons and single-particle excitations of the 2DEG is revealed, with the lower hybrid mode being damped when its dispersion overlaps with the 2DEG single-particle continuum.

Dispersion of the lower hybrid mode softens to develop the roton minimum which becomes deeper with the increase of the polariton density $n_\mathrm{p}$. It is caused by the effective polariton-polariton attraction mediated by the density response of the 2DEG, as also noted in \cite{Cotlet,Shelykh}. Above the critical polariton density $n_\mathrm{p}^\mathrm{crit}$, the system becomes unstable against supersolid \cite{Bland,Hertkorn,supersolid} or density wave formation, since the energy of the roton minimum becomes negative. As specific physical realizations, we considered the setups based on two-dimensional TMD bilayers and coupled semiconductor QWs. In both cases we consider the spatially indirect excitons which interact with the neighbouring 2DEG relatively strongly due to their persistent dipole moment. For the TMD-based setup, the instability occurs at $n_\mathrm{p}^{\rm crit}\sim10^{12}-10^{13}\,\mbox{cm}^{-2}$, which could be close to the experimentally relevant conditions. In contrast, the setup based of GaAs-type QWs requires high polariton densities $n_\mathrm{p}^{\rm crit}\sim10^{11}\,\mbox{cm}^{-2}$, which are much harder to be achieved due to larger excitonic Bohr radius in these materials.

In a close vicinity of the critical polariton density $n_\mathrm{p}^\mathrm{crit}$, the low-energy rotons dominate the pairing interaction, as evidenced by the analysis of their Eliashberg functions in Sec.~\ref{Sec_Eliashberg_func}, and this regime is the most promising for a possible superconducting pairing. We analyze superconducting Cooper pairing in the 2DEG induced by the screened electron-electron interaction $V_\mathrm{ee}$ in different approximations. The first one is the numerical solution of the set of self-consistency equations for momentum and frequency dependent gap and renormalization functions. Direct solution of these equations provides $T_\mathrm{c}$ up to 0.1 K and 0.03 K for, respectively, TMD- and QW-based setups with the screening in TFA. Taking into account dynamical screening in RPA provides controversial results: in the TMD-based case, due to entering the strongly-interacting regime, the unphysical Cooper pairing in the 2DEG appears even in the absence of polaritons. This artefact of RPA \cite{Rietschel} can be avoided by a more accurate treatment of the interaction screening when taking into account vertex corrections \cite{Takada, Richardson}. This is not the case for the QW-based system, where the electron interactions are not as strong, and RPA results in maximal $T_\mathrm{c}\sim0.01\,\mbox{K}$.

Another widely used approach that we employed to calculate $T_\mathrm{c}$ is the Eliashberg equations, where the momentum dependence of the gap and renormalization functions is neglected, and only the frequency dependence is retained. Taking realistic values of the Coulomb pseudopotential which characterizes the effective electron-electron repulsion competing with the polariton-induced attraction, we obtain $T_\mathrm{c}$ close to those found from the aforementioned full numerical solution in TFA. The Allen-Dynes formula, which approximates solutions of the Eliashberg equations, also provides the similar results. Nevertheless, the assumption of smooth momentum dependence of the gap and renormalization functions is not justified in our case (see Appendix~\ref{Appendix_B}), so care should be taken when transferring the well-established methods of the phonon-mediated superconductivity to other pairing mechanisms. The third approach, namely the BCS approximation, where the dynamical effects are neglected, results in extremely high $T_\mathrm{c}$ reaching room temperatures, in agreement with the results of other authors \cite{Laussy_PRL,Laussy_JNP,Cherotchenko,Sedov,Sun_2DMater,Sun_Thesis,Sun_NJP}. Appendix~\ref{Appendix_C} is devoted to the origin of such a discrepancy between the Eliashberg and BCS approaches. In the BCS approach, the mapping of characteristic energy width of the pairing region (determined by the roton energies in the vicinity of the instability) to the momentum space of electrons distorts the results due to significant mismatch between the electron and bogolon dispersions. As a consequence, the energy width of the pairing region in the static BCS approach is determined by the energy of electron excitations possessing the roton momentum, which is 2-3 orders of magnitude larger than the original energy of rotons. The latter is taken into account in the dynamical Eliashberg approach. This results in the corresponding difference of critical temperatures predicted in the two approaches. We conclude that the predictions of room-temperature superconductivity induced by the polariton-BEC mechanism obtained in the BCS approach are highly biased due to the neglection of dynamical effects. Other calculations based on the Eliashberg approach that appear in the literature~\cite{Sun_PRR,Julku}, which also predicted high $T_\mathrm{c}$ about tens and hundreds of Kelvins, correctly took into account dynamical effects but neglected the screening of interaction by the 2DEG.

According to our calculations which take into account both the screening and dynamical effects, $T_\mathrm{c}$ could reach no more than fractions of a Kelvin, and only in a very close vicinity of the roton instability. For completeness of our analysis, we also studied the role of the pair-bogolon, or noncondensate, processes which were recently proposed in Refs.~\cite{Sun_2DMater,Sun_Thesis,Sun_NJP,Sun_PRR} to be dominant in the pairing interaction. As shown in Appendix~\ref{Appendix_D}, when the interaction screening is taken into account, the contribution of pair-bogolon processes to the electron-electron pairing interaction appears to be very small.

The approximate analytical expressions presented in Appendix~\ref{Appendix_A} allow to estimate the instability threshold $n_\mathrm{p}^\mathrm{crit}$ for polariton density, superconducting coupling constant $\lambda$, and critical temperature $T_\mathrm{c}$ at given system parameters. We also considered the possibility of the $(p_x+ip_y)$-wave pairing suggested for the Bose-Fermi systems \cite{Cotlet,Julku}. In this case the pairing interaction $V_\mathrm{ee}$ in Eqs.~(\ref{Z_eq})--(\ref{phi_eq}) should be multiplied by the angular factor $\cos(\widehat{\mathbf{k},\mathbf{k}^\prime})$. It does not change appreciably the coupling constant $\lambda$ because of the dominating contribution of the roton minimum at rather low momentum $q_0\ll k_\mathrm{F}$, although the reduction of the Coulomb repulsion (and hence the Coulomb pseudopotential $\mu$) is stronger in the $p$-wave channel. Our numerical calculations with taking into account both these effects show that $T_\mathrm{c}$ for the $p$-wave pairing can be slightly higher than in the case of ordinary $s$-wave pairing at the same conditions.

The main conclusion of this paper is that interaction screening and dynamical effects are both crucial for correct analysis of low-temperature many-body phenomena in the Bose-Fermi hybrid systems of strongly coupled exciton-polaritons and electrons. Possible directions of future research may include analysis of polaronic effects at lower exciton densities, experimental detection of the predicted hybrid upper and lower modes, and analysis of enhancement of preexisting superconductivity due to polariton BEC in a microcavity \cite{Skopelitis}. Fermi-polaron effects, i.e. dressing of individual polaritons by a cloud of virtual excitations of 2DEG (which reduces to the trion formation at low electron density), according to the recent studies, can renormalize and split the polariton energies \cite{Imamoglu1,Bastarrachea-Magnani,Zhumagulov}, induce additional 2DEG-mediated polariton-polariton interactions \cite{Imamoglu2,Muir}, and affect polariton BEC \cite{Julku2021}. From the point of view of our diagrammatic approach, the polaronic effects correspond to ladder diagrams where the interlayer electron-exciton interaction connect the Green's functions of electrons and bogolons multiple times. Such processes can be referred to as vertex corrections to the pairing interaction, and lie beyond our mean-field approach. Moreover, the ladder diagrams, involving the scattering of bogolons, are of the non-condensate type and thus do not get the benefit from Bose stimulation, which enhances the mean-field pairing interaction [see Fig.~\ref{Fig_Diag}(d,e)]. Nevertheless, such processes can be enhanced in a close vicinity of the roton instability due to low energy of bogolons with the roton momenta, as well as other correlation effects.

As a further outlook, going beyond the mean-field approximations used in our calculations could allow to possibly discover new regimes of the considered electron-polariton system in the vicinity of (or inside) the supersolid phase. For example, the non-perturbative dynamical mean-field theory (DMFT) revealed an interplay of dynamical phonon-mediated and static Coulomb interactions \cite{Bauer} in electron Cooper pairing. At the same time, recently the nonlocal extensions of DMFT taking into account both frequency and momentum dependencies have significantly advanced the theory of superconductivity in correlated electron systems \cite{Rohringer,Kitatani}. Therefore development of similar non-perturbative approaches for coupled Bose-Fermi systems could be very promising to support the experimental progress in the area of microcavity-integrated 2D materials. Treating excitons and photons as separate particles subject to intertwined Bose-condensations and at the same time interacting with the 2DEG can also reveal novel phases \cite{Strashko}.

\section*{Acknowledgments}
The work was supported by the Russian Foundation for Basic Research (RFBR) within the Project No. 21–52–12038. N.S.V. acknowledges the financial support of the NRNU MEPhI Priority 2030 program. Yu.E.L. thanks Foundation for the advancement of theoretical physics and mathematics ``Basis''.

\appendix

\section{Analytical estimates}\label{Appendix_A}

When the polariton density $n_\mathrm{p}$ is close to the critical one $n_\mathrm{p}^\mathrm{crit}$, the major contribution to the coupling constant $\lambda$ comes from the vicinity of low-energy roton minimum. In this Appendix, we provide approximate formulae which could be useful for the estimation of $n_\mathrm{p}^{\rm crit}$, $\lambda$, and $T_\mathrm{c}$ in this regime.

Since the roton minimum is very low in frequency, we can resort to the TFA static limit of (\ref{Bog_disp}):
\begin{equation}
E_q^\mathrm{p}=\left.\sqrt{\tilde\epsilon_q^\mathrm{p}(\tilde\epsilon_q^\mathrm{p}+2X_0^2X_q^2n_0^\mathrm{p}\tilde{v}_q^\mathrm{xx})}\right|_{\Pi_\mathrm{e}\rightarrow\Pi_\mathrm{e}^\mathrm{TFA}}.\label{E_q_TFA}
\end{equation}
To find $n_\mathrm{p}^{\rm crit}$, we consider the condition when (\ref{E_q_TFA}) touches the abscissa at $q=q_0$:
\begin{equation}
E_q^\mathrm{p}|_{q=q_0}=0,\quad \partial E_q^\mathrm{p}/\partial q|_{q=q_0}=0.
\end{equation}
As seen in Fig.~\ref{Fig_disp}, at $q=q_0$ one is deeply in the excitonic part of the polariton dispersion (dashed lines), so we can take $\epsilon_q^\mathrm{p}\approx(\sqrt{\delta^2+4\Omega_\mathrm{R}^2}-\delta)/2=(1-X_0^2)\sqrt{\delta^2+4\Omega_\mathrm{R}^2}$,  $X_q^2\approx1$, and
\begin{equation}
X_0^2\approx\frac12\left(1+\frac\delta{\sqrt{\delta^2+4\Omega_\mathrm{R}^2}}\right).\label{X_0sq}
\end{equation}
Introducing the function $f_{n_\mathrm{p}}(q)=\tilde\epsilon_q^\mathrm{p}+2X_0^2X_q^2n_\mathrm{p}\tilde{v}_q^\mathrm{xx}$, we rewrite (\ref{E_q_TFA}) in the form
\begin{equation}
E_q^\mathrm{p}=\sqrt{\tilde\epsilon_q^\mathrm{p}f_{n_\mathrm{p}}(q)}.\label{Bog_disp_approx}
\end{equation}
Noting that $q_0d,q_0a_\mathrm{B}\sim0.2\ll1$, we can simplify the electron-exciton interaction (\ref{v_ex}) as $v_q^\mathrm{ex}\approx2\pi e^2de^{-qL}/\varepsilon$, and, using (\ref{tilde_v_xx}), we obtain in TFA
\begin{align}
f_{n_\mathrm{p}}(q)&\approx(1-X_0^2)\sqrt{\delta^2+4\Omega_\mathrm{R}^2}\nonumber\\&+2n_\mathrm{p}X_0^2\left\{U-\frac{2\pi e^2d^2e^{-2qL}}\varepsilon\frac{qq_\mathrm{TF}}{q+q_\mathrm{TF}}\right\},\label{f}
\end{align}
where $q_\mathrm{TF}=ge^2m_\mathrm{e}^*/\varepsilon$ is the Thomas-Fermi screening wave vector. From the roton minimum condition $f'_{n_\mathrm{p}}(q_0)\approx0$ we find $q_0^2+q_0q_\mathrm{TF}-q_\mathrm{TF}/2L\approx0$, and, since $2q_\mathrm{TF}L\gg1$, we obtain the approximate momentum of the roton minimum
\begin{equation}
q_0\approx1/2L.\label{q_0}
\end{equation}
Substituting (\ref{q_0}) to (\ref{f}) and equating $f_{n_\mathrm{p}^\mathrm{crit}}(q_0)=0$ we obtain the critical density of polaritons
\begin{equation}
n_\mathrm{p}^\mathrm{crit}=\frac{(1-X_0^2)\sqrt{\delta^2+4\Omega_\mathrm{R}^2}}{2X_0^2\left\{\pi e^2d^2/\varepsilon L\exp(1)-U\right\}}.\label{n_c}
\end{equation}
The formula (\ref{n_c}) can be used together with (\ref{X_0sq}) to estimate an order of magnitude of $n_\mathrm{p}^\mathrm{crit}$ at given system parameters. However the exact value of $n_\mathrm{p}^\mathrm{crit}$ which defines the phase boundary in Fig.~\ref{Fig_phase} may be several times higher or lower due to a more complex momentum dependence of the Bogoliubov dispersion and electron-exciton interaction.

When the polariton density $n_\mathrm{p}$ considered in our calculations is slightly lower than the critical one, $n_\mathrm{p}=n_\mathrm{p}^\mathrm{crit}-\delta n_\mathrm{p}$, $\delta n_\mathrm{p}\ll n_\mathrm{p}^\mathrm{crit}$, we can approximate $f_{n_\mathrm{p}}(q)\approx f_{n_\mathrm{p}}(q_0)+\frac12(q-q_0)^2f''_{n_\mathrm{p}}(q_0)$ near the roton minimum. Using (\ref{f}), one finds $f_{n_\mathrm{p}}(q_0)=(\delta n_\mathrm{p}/n_\mathrm{p}^\mathrm{crit})(1-X_0^2)\sqrt{\delta^2+4\Omega_\mathrm{R}^2}$, $f''_{n_\mathrm{p}}(q_0)\approx4L^2(1-X_0^2)\sqrt{\delta^2+4\Omega_\mathrm{R}^2}$. In the latter formula we have taken $f''_{n_\mathrm{p}}(q_0)\approx f''_{n_\mathrm{p}^\mathrm{crit}}(q_0)$ and used (\ref{n_c}) with neglecting $U$ in the denominator, which is sufficient for a rough estimate of the steepness of $f_{n_\mathrm{p}}(q)$ near the minimum. The coupling constant $\lambda$ can be calculated in RPA (\ref{lambda_2}) or in TFA (\ref{lambda}), because these expressions both provide close results. In the latter case, the polariton-induced electron attraction (\ref{V_2}) reads $V_2(q,i\omega_n)=2s_q^2E_q^\mathrm{p}/[(i\omega_n)^2-(E_q^\mathrm{p})^2]$. It is mediated by the exchange of Bogoliubov quasiparticles with the squared interaction vertex
\begin{equation}
s_q^2=\frac{X_0^2n_\mathrm{p}\tilde\epsilon_q^\mathrm{p}}{E_q^\mathrm{p}}\left[\frac{2\pi e^2qde^{-qL}}{\varepsilon(q+q_\mathrm{TF})}\right]^2.
\end{equation}
The Eliashberg function (\ref{alpha2F_2}) for this kind of interaction is $\alpha^2F_2(\omega)=\mathcal{N}\langle s_{|\mathbf{k}-\mathbf{k}'|}^2\delta(\omega-E_{\mathbf{k}-\mathbf{k}'}^\mathrm{p})\rangle_\mathrm{FS}$, and the coupling constant (\ref{lambda_2}) is
\begin{equation}
\lambda=\frac{2\mathcal{N}}\pi\int_0^\pi\frac{s_q^2}{E_q^\mathrm{p}}\,d\varphi,\quad q=2k_\mathrm{F}\sin\frac\varphi2.\label{lambda_int}
\end{equation}
The main contribution to the integral (\ref{lambda_int}) comes from the vicinity of the roton minimum $q_0$. Assuming $q_0\ll k_\mathrm{F}$ and using (\ref{Bog_disp_approx}) with the quadratic decomposition of $f_{n_\mathrm{p}}(q)$ introduced above, we obtain the estimate of the coupling constant at given system parameters:
\begin{align}
\lambda&=\frac1{\sqrt{1-U\varepsilon L\exp(1)/\pi e^2d^2}}\nonumber\\ &\times\frac{\varepsilon}{4\sqrt{2\pi n_\mathrm{e}}g^{3/2}m_\mathrm{e}^*e^2L^2\sqrt{\delta n_\mathrm{p}/n_\mathrm{p}^\mathrm{crit}}}.\label{lambda_est}
\end{align}
This formula is relatively simple and provides a correct order of magnitude for $\lambda$, as well as its dependence on values of system parameters. Note that $\lambda$ diverges when the polariton density approaches the critical value, i.e. $\delta n_\mathrm{p}=n_\mathrm{p}^\mathrm{crit}-n_\mathrm{p}\rightarrow0$.

When estimating the critical temperature $T_\mathrm{c}$, we note that it does not exceed the characteristic energy of Bogoliubov quasiparticles near the roton minimum $\langle\omega\rangle=\sqrt{\tilde\epsilon_{q_0}^\mathrm{p}f_{n_\mathrm{p}}(q_0)}$. Using (\ref{f}), we obtain:
\begin{equation}
T_\mathrm{c}\lesssim\langle\omega\rangle=\sqrt{\frac{\delta n_\mathrm{p}}{n_\mathrm{p}^\mathrm{crit}}}\frac{(1-X_0^2)\sqrt{\delta^2+4\Omega_\mathrm{R}^2}}{\sqrt{1-U\varepsilon L\exp(1)/\pi e^2d^2}}.\label{omega_approx}
\end{equation}
This formula provides a correct order of magnitude of maximal achievable $T_\mathrm{c}$ in the vicinity of the roton instability $\delta n_\mathrm{p}\rightarrow0$.

\section{Frequency and momentum dependence of the functions $Z$ and $\varphi$}\label{Appendix_B}

Here we analyze how the functions $Z$ and $\varphi$ found from the self-consistency equations (\ref{Z_eq})--(\ref{phi_eq}) depend on momentum ${\bf k}$ and the Matsubara frequency $\varepsilon_n$. In the Eliashberg approach \cite{Eliashberg} their momentum dependence is assumed to be smooth enough to take $Z(\mathbf{k},i\varepsilon_n)\approx Z(k_\mathrm{F},i\varepsilon_n)$, $\varphi(\mathbf{k},i\varepsilon_n)\approx \varphi(k_\mathrm{F},i\varepsilon_n)$, while in the BCS approach \cite{Laussy_JNP}, oppositely, the frequency dependence is neglected: $\varphi(\mathbf{k},i\varepsilon_n)\approx\varphi(\mathbf{k})$. Our goal is to check whether these assumptions are justified in our case.

\begin{figure}[b]
\begin{center}
\includegraphics[width=\columnwidth]{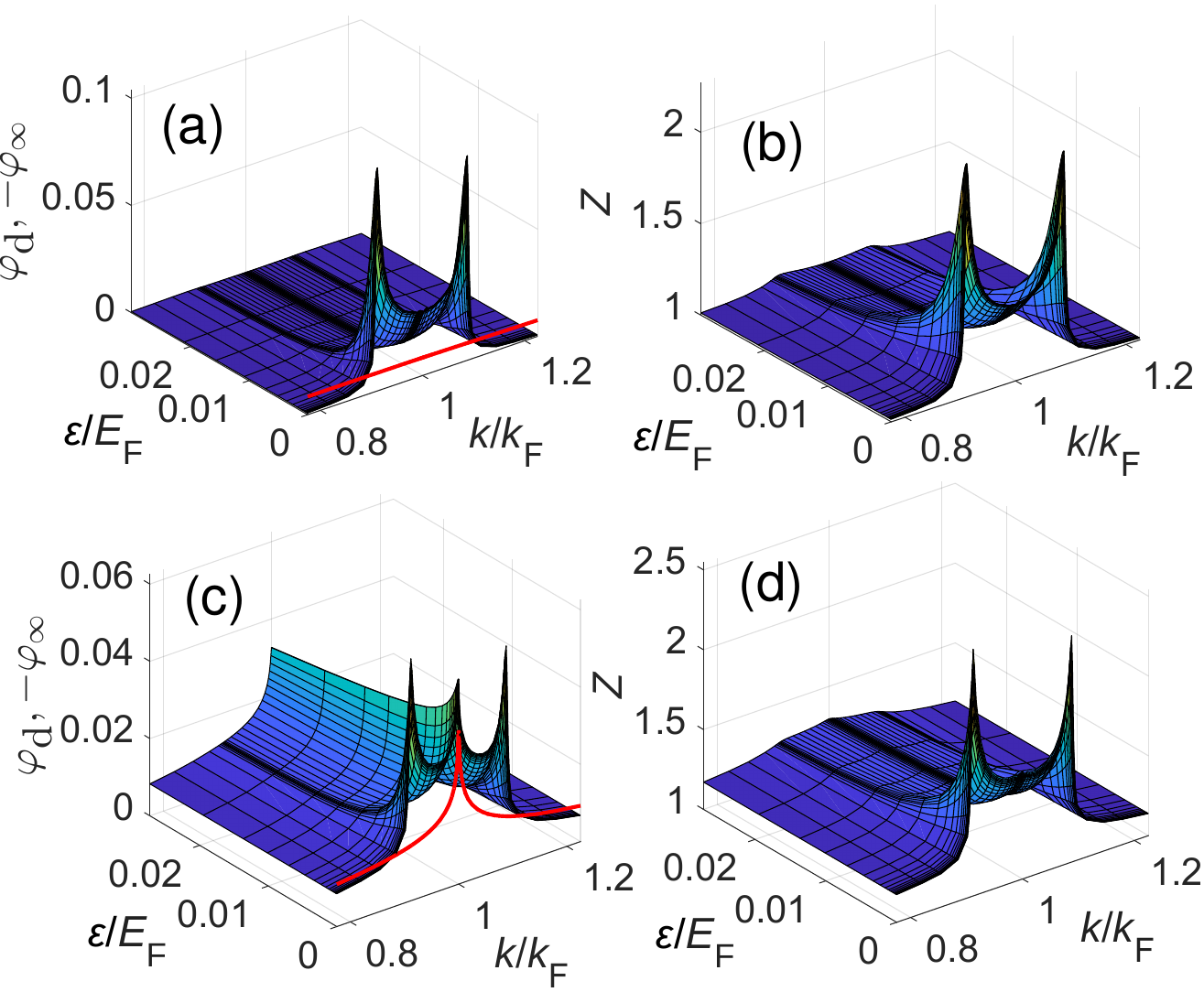}
\end{center}
\caption{\label{Fig_Zphi}(a,c) Frequency and momentum dependence of $-\varphi_\infty$ (red solid lines) and $\varphi_\mathrm{d}$ and (b,d) the same for $Z$, for the QW-based system at the polariton density $n_\mathrm{p}=3.103\times10^{11}\,\mbox{cm}^{-2}$}. Top (a,b) and bottom (c,d) panels correspond to the solution of the self-consistency equations (\ref{Z_eq})--(\ref{phi_eq}) with the interaction $V_\mathrm{ee}$ taken in TFA and RPA, respectively.
\end{figure}

Similarly to the separation of anomalous self-energy into ``Coulomb'' and ``phonon'' parts in the Eliashberg theory \cite{Scalapino}, we represent $\varphi(\mathbf{k},i\varepsilon_n)=\varphi_\infty(k)+\varphi_\mathrm{d}(k,i\varepsilon_n)$ as a sum of the frequency-independent $\varphi_\infty(k)=\lim_{\varepsilon_n\rightarrow\infty}\varphi(k,i\varepsilon_n)$ and the remaining dynamical $\varphi_\mathrm{d}(k,i\varepsilon_n)$ parts. In Fig.~\ref{Fig_Zphi}(a,b) we plot $\varphi_\infty$, $\varphi_\mathrm{d}$, and $Z$ as functions of $k$ and Matsubara frequency $\varepsilon$ for the case of the QW-based system when the interaction $V_\mathrm{ee}$ is screened in TFA [which corresponds to the solid line in Fig.~\ref{Fig_Tc}(d)]. Note that the function $\varphi_\infty$ depends only on $k$, so it is plotted by the red curve. For convenience, we plot $\varphi_\infty$ with the negative sign, so that the total gap function $\varphi$ is represented by the difference between the 2D surface and 1D line in Fig.~\ref{Fig_Zphi}(a).

Both $\varphi_\mathrm{d}$ and $Z$ display sharp peaks at momenta $k=k_\mathrm{F}\pm q_0$, where $q_0$ is the momentum corresponding to the roton minimum [see Fig.~\ref{Fig_disp}(a)]. This reflects the important role of low-frequency excitations (at the finite momentum $q_0$) that amplify the pairing strength. We see that the assumption of smoothness of the momentum dependence, which could be plausible for short-range electron-phonon interaction in metals, is barely justified in the case of polariton-mediated superconductivity, especially since the peak values are several times larger than those at the Fermi surface. Nevertheless, the Eliashberg approach and full solution of the self-consistency equations eventually provide similar results for $T_\mathrm{c}$ [compare solid and dashed lines in Fig.~\ref{Fig_Tc}(d)]. Note that $\varphi_\infty(k)$ [red line in Fig.~\ref{Fig_Zphi}(a)] remains smooth due to the short-range character of the TFA-screened Coulomb interaction $V_1$.

Figs.~\ref{Fig_Zphi}(c,d) show what happens when the interaction $V_\mathrm{ee}$ is dynamically screened in RPA [which corresponds to Fig.~\ref{Fig_Tc}(e)]. The peaks in $\varphi_\mathrm{d}$ and $Z$ at  $k=k_\mathrm{F}\pm q_0$ become even sharper. In addition, the new peak develops at $k=k_\mathrm{F}$ in both $\varphi_\infty$ and $\varphi_\mathrm{d}$, which reflects the long-range singularity of Coulomb interaction being averaged over the Fermi surface. However, the peaks in $\varphi_\mathrm{d}$ and $\varphi_\infty$ at $k=k_\mathrm{F}$ almost compensate each other, which demonstrates the cancellation of divergences due to direct Coulomb repulsion $v_q^\mathrm{ee}$ and to plasmon-mediated attraction in $V_1$ discussed in Sec.~\ref{Sec_Eliashberg_func}. Thus the assumption of momentum smoothness of $\varphi$ and $Z$ near the Fermi surface, which is crucial for derivation of Eliashberg equations, becomes even more unrealistic when screening is accounted for in RPA.

\section{Comparison of the Eliashberg and BCS approaches}\label{Appendix_C}

In this Appendix, we investigate the origin of the orders-of-magnitude difference in the critical temperatures obtained in the Eliashberg (see Sec.~\ref{Sec_Eliashberg}) and BCS (Sec.~\ref{Sec_BCS}) approaches. We consider the simplified version of the self-consistent gap equation (\ref{phi_eq}) from which both the Eliashberg and BCS gap equations are derived:
\begin{equation}
\varphi(\mathbf{k},i\varepsilon_n)=-\frac{T}S\sum_{\mathbf{k}'\varepsilon_m}V_2(|\mathbf{k}-\mathbf{k}'|,i\varepsilon_n-i\varepsilon_m)\frac{\varphi(\mathbf{k}',i\varepsilon_m)}{\varepsilon_m^2+\xi_{k'}^2}.\label{gap_eq1}
\end{equation}
Here we omitted the renormalization function $Z$ and the part $V_1$ of the total electron-electron interaction potential $V_\mathrm{ee}=V_1+V_2$ which corresponds to the screened electron repulsion. Performing the integration over direction of $\mathbf{k}'$ and switching from momenta $k$, $k'$ to the electron energies counted from the Fermi level $\xi=k^2/2m_\mathrm{e}^* - E_\mathrm{F}$, $\xi'=k'^2/2m_\mathrm{e}^* - E_\mathrm{F}$, we obtain
\begin{equation}
\varphi(\xi,i\varepsilon_n)=T\sum_{\varepsilon_m}\int d\xi'\:F(\xi,\xi',\varepsilon_n-\varepsilon_m)\frac{\varphi(\xi',i\varepsilon_m)}{\varepsilon_m^2+\xi'^2},\label{gap_eq2}
\end{equation}
where the dimensionless pairing attraction is
\begin{equation}
F(\xi,\xi',\omega)=-\frac{\mathcal{N}}{2\pi}\int_0^{2\pi}d\theta\:V_2(|\mathbf{k}-\mathbf{k}'|,i\omega)
\label{pairing_int}
\end{equation}
($k$ and $k'$ are related to $\xi$ and $\xi'$, and $\theta$ is the angle between the vectors $\mathbf{k}$ and $\mathbf{k}'$).

Specifically, here we consider the QW-based system at the polariton density $n_\mathrm{p}=3.103\times10^{11}\,\mbox{cm}^{-2}$, where the Eliashberg approach predicts $T_\mathrm{c}=5.6\times10^{-4}\,\mbox{K}$, while the BCS approach predicts $T_\mathrm{c}=0.38\,\mbox{K}$ which is 3 orders of magnitude higher [cf. Figs.~\ref{Fig_Tc}(d) and \ref{Fig_Tc}(f)]. For simplicity, we will consider the interaction screening in TFA where the renormalization function $Z\sim1.5$ in the most relevant region near the Fermi surface [see Fig.~\ref{Fig_Zphi}(b)], and the screened electron-electron repulsion $V_1(q)=2\pi e^2/\varepsilon(q+q_\mathrm{TF})$, being multiplied by the density of states at the Fermi level $\mathcal{N}=m_\mathrm{e}^*/2\pi$, provides the contribution $\langle\mathcal{N}V_1(q)\rangle_\mathrm{FS}=0.12$ at the Fermi surface, which is significantly smaller than $F$ near the Fermi level. Thus the simplifications in Eq.~(\ref{gap_eq2}) may bring some quantitative errors to the calculated $T_\mathrm{c}$, but they should not obscure the origin of the large quantitative discrepancy between the two predictions.

\begin{figure}[t]
\begin{center}
\includegraphics[width=\columnwidth]{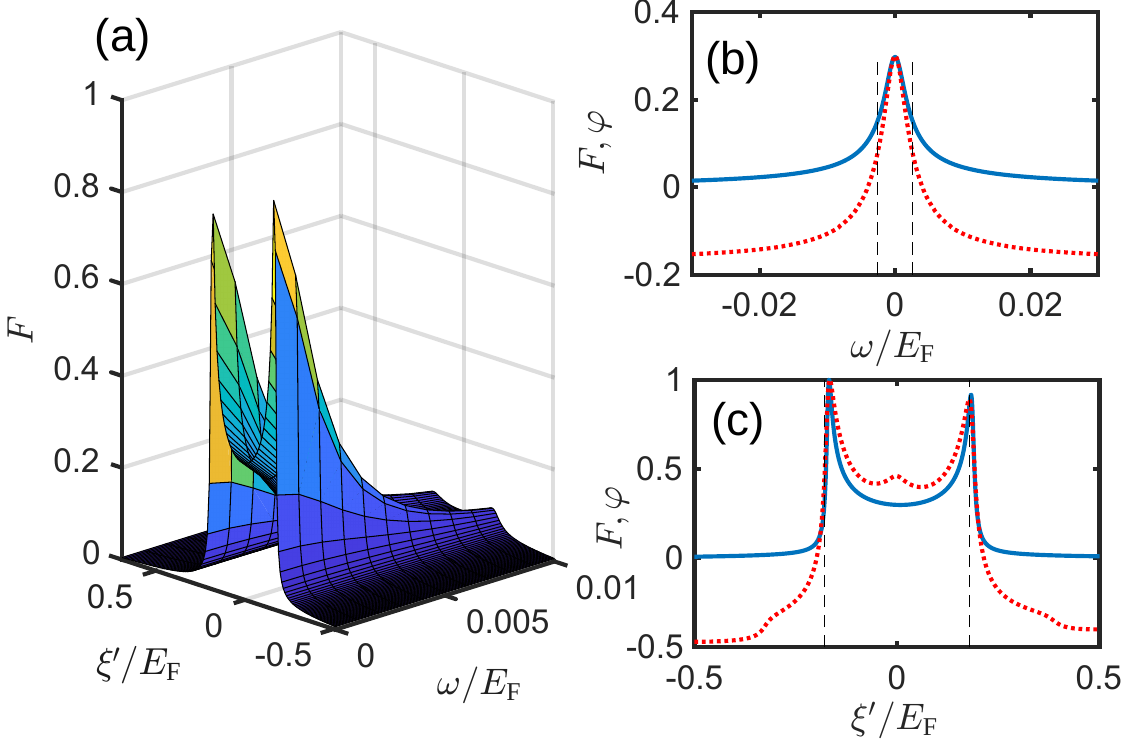}
\end{center}
\caption{\label{Fig_El_BCS}(a) Dimensionless pairing interaction $F(0,\xi',\omega)$ as a function of the energy distances from the Fermi level along momentum $\xi'$ and frequency $\omega$ axes calculated for the QW-based system. (b) Cross section $F(0,0,\omega)$ along the frequency axis used in the Eliashberg approach (solid line) plotted together with the gap function $\varphi(i\omega)$ obtained by numerical solution of the Eliashberg equations (\ref{Z_Eliashberg_reg})--(\ref{phi_Eliashberg_reg}) (dotted line, in arbitrary units). (c) Cross section $F(0,\xi',0)$ along the momentum axis used in the BCS approach (solid line) plotted together with the gap function $\varphi(\xi')$ obtained numerically from the BCS gap equation (\ref{BCS}) (dotted line, in arbitrary units). Vertical dashed lines in (b) and (c) show characteristic widths of both the pairing interaction and the gap functions along the frequency ($|\omega|\lesssim aE_0$) and momentum ($|\xi'|<\xi_0$) axes, respectively.}
\end{figure}

Physically, the pairing interaction $F(\xi,\xi',\omega)$ is dominated by the contribution of virtual bogolons near the roton minimum [see Fig.~\ref{Fig_disp}(b)] at the momentum $q_0=0.022\,\mbox{nm}^{-1}$ and energy $E_0\equiv E_{q_0}^\mathrm{p}=0.054\,\mbox{meV}$. Hence it is approximately confined within the region $k,k'>q_0$, $|k-k'|<q_0$ in momentum space, and has the width of the order of $E_0$ along the frequency axis. Proceeding further towards approximate solutions of the gap equation (\ref{gap_eq2}), in the l.h.s. we fix both the momentum and energy at the Fermi surface ($\xi=0$, $\varepsilon_n=0$), while for the r.h.s we assume that $\varphi$ is approximately constant in the integral throughout momentum and frequency ranges where $F$ provides the dominating contribution. Such approximations, which are widely used in both the Eliashberg and BCS approaches to estimate $T_\mathrm{c}$, yield the equation
\begin{equation}
1=T\sum_{\varepsilon_m}\int d\xi'\:F(0,\xi',\varepsilon_m)\frac1{\varepsilon_m^2+\xi'^2}.\label{gap_eq3}
\end{equation}
Fig.~\ref{Fig_El_BCS}(a) showing the function $F(0,\xi',\omega)$ confirm our expectations: the interaction is approximately confined to the region $|\xi'|<\xi_0\equiv k_\mathrm{F}q_0/m_\mathrm{e}^*=0.18E_\mathrm{F}$ in momentum space and has the half width at half maximum $\omega\lesssim aE_0= 0.0026E_\mathrm{F}$ in the frequency space (where the factor $a\approx1.72$ is introduced to account for the effective increase of the characteristic roton energy with respect to the minimal $E_0$ due to momentum integration). We can crudely assume the following model form of this function:
\begin{equation}
F(0,\xi',\omega)\approx C\frac{\Theta(\xi_0-|\xi'|)}{(\omega/aE_0)^2+1},\label{pairing_int_appr}
\end{equation}
with the typical magnitude near the Fermi level $C\approx0.3$.

In the Eliashberg approach (see Sec.~\ref{Sec_Eliashberg}), we neglect the momentum (i.e. $\xi'$) dependence of $F$ and replace $F(0,\xi',\omega)$ by $F(0,0,\omega)\approx C/[(\omega/aE_0)^2+1]$. Fig.~\ref{Fig_El_BCS}(b) demonstrates that $F(0,0,\omega)$ has indeed the characteristic frequency width $aE_0$. The actual gap function $\varphi(i\omega)$ obtained numerically from the complete Eliashberg equations also demonstrates the confinement of its rapidly changing part to this region (apart from slowly decaying tails originating from Coulomb interaction). After performing the integration over $\xi'$ in the gap equation (\ref{gap_eq3}), we obtain
\begin{equation}
1=\pi T\sum_{\varepsilon_m}\frac{C}{|\varepsilon_m|\{(\varepsilon_m/aE_0)^2+1\}}.
\end{equation}
Summation over the fermionic Matsubara frequencies $\varepsilon_m=\pi T(2m+1)$ can be performed analytically in the limit $T\ll E_0$ in terms of the digamma function \cite{Marsiglio}, so the resulting estimate for $T_\mathrm{c}$ in the Eliashberg approach is
\begin{equation}
T_\mathrm{c}^\mathrm{El}=(2e^\gamma/\pi)aE_0e^{-1/C},\label{Tc_El}
\end{equation}
where $\gamma\approx0.577$ is the Euler's gamma constant.

In the BCS approach (see Sec.~\ref{Sec_BCS}), we neglect the frequency dependence of $F$ and replace $F(0,\xi',\omega)$ by $F(0,\xi',0)\approx C\Theta(\xi_0-|\xi'|)$. Fig.~\ref{Fig_El_BCS}(c) confirms the approximate confinement of both the static interaction $F(0,\xi',0)$ and numerical solution $\varphi(\xi')$ of the BCS gap equation (again, apart from the slowly decaying tails) to the region $|\xi'|<\xi_0$. After summation over $\varepsilon_m$, Eq.~(\ref{gap_eq3}) reduces to the simplified BCS equation for critical temperature
\begin{equation}
1=\int_{-\xi_0}^{\xi_0}d\xi'\:\frac{C}{2|\xi'|}\tanh\frac{|\xi'|}{2T}.
\end{equation}
The standard solution of this equation in the limit $T~\!\!\ll~\!\!\xi_0$ provides in turn the BCS estimate
\begin{equation}
T_\mathrm{c}^\mathrm{BCS}=(2e^\gamma/\pi)\xi_0e^{-1/C}.\label{Tc_BCS}
\end{equation}

Our estimates (\ref{Tc_El}) and (\ref{Tc_BCS}) for the critical temperature in both approaches provide $T_\mathrm{c}^\mathrm{El}\approx0.044\,\mbox{K}$ and $T_\mathrm{c}^\mathrm{BCS}\approx2.98\,\mbox{K}$ (given that $E_\mathrm{F}=415\,\mbox{K}$). These values are overestimated in comparison to $T_\mathrm{c}^\mathrm{El}=5.6\times10^{-4}\,\mbox{K}$ and $T_\mathrm{c}^\mathrm{BCS}=0.38\,\mbox{K}$ obtained by numerical solutions of the Eliashberg and BCS equations, respectively [see Figs.~\ref{Fig_Tc}(d,f)], because we have used several crude approximations to obtain the above analytical results. Nevertheless, the presented analysis demonstrates the key difference between the Eliashberg and BCS approaches: in the former, the pairing region is restricted by the characteristic energy $E_0$ of \emph{bogolons} near the roton minimum, while in the latter, this region is determined by the energy $\xi_0\approx\epsilon_{k_\mathrm{F}+q_0}^\mathrm{e}-E_\mathrm{F}$ acquired by \emph{electrons} on the Fermi surface when they emit or absorb bogolons with the roton-minimum momentum $q_0$. Since the roton minimum in our case is located far below the edge of electron-hole continuum [see Fig.~\ref{Fig_disp}(b)], $\xi_0$ is 2-3 orders of magnitude higher than $E_0$, which results in the corresponding huge difference between $T_\mathrm{c}^\mathrm{BCS}$ and $T_\mathrm{c}^\mathrm{El}$. The same feature is demonstrated by numerical solutions of the Eliashberg and BCS equations obtained without the simplifications of this Appendix: the characteristic energy width of both the pairing region and the gap function in the BCS approach [Fig.~\ref{Fig_El_BCS}(c)] is 2-3 orders of magnitude larger than in the Eliashberg approach [Fig.~\ref{Fig_El_BCS}(b)], which results in the proportional ratio of the critical temperatures  obtained in these calculations.

\section{Role of noncondensate processes}\label{Appendix_D}

In this section of Appendix we estimate the contribution of noncondensate, or ``pair-bogolon'', processes to the excitonic density response function $\Pi_\mathrm{x}$, which were claimed to be dominant in the pairing interaction $V_\mathrm{ee}$ and superconducting coupling constant \cite{Sun_2DMater,Sun_Thesis,Sun_NJP,Sun_PRR}. Consideration of these processes results in an additional term, $\Pi_\mathrm{x}\rightarrow\Pi_\mathrm{x}+\Pi^\mathrm{nc}_\mathrm{x}$, in the excitonic density response beyond the dominating \cite{Griffin} condensate contribution (\ref{Pi_x}). The noncondensate density response
\begin{multline}
\Pi^\mathrm{nc}_\mathrm{x}(q,i\omega_n)=-\frac1S\sum_{\mathbf{k}}X_k^2X_{|\mathbf{k}-\mathbf{q}|}^2\\
\times\left\{(u_{\mathbf{k}}u_{\mathbf{k}-\mathbf{q}}+v_{\mathbf{k}}v_{\mathbf{k}-\mathbf{q}})^2\frac{(f_{\mathbf{k}}-f_{\mathbf{k}-\mathbf{q}})(E^\mathrm{p}_{\mathbf{k}}-E^\mathrm{p}_{\mathbf{k}-\mathbf{q}})}{(i\omega_n)^2-(E^\mathrm{p}_{\mathbf{k}}-E^\mathrm{p}_{\mathbf{k}-\mathbf{q}})^2}\right.\\
\left.-(u_{\mathbf{k}}v_{\mathbf{k}- \mathbf{q}}+v_{\mathbf{k}}u_{\mathbf{k}- \mathbf{q}})^2\frac{(1+f_{\mathbf{k}}+f_{\mathbf{k} -\mathbf{q}})(E^\mathrm{p}_{\mathbf{k}}+E^\mathrm{p}_{\mathbf{k}-\mathbf{q}})}{(i\omega_n)^2-(E^\mathrm{p}_{\mathbf{k}} +E^\mathrm{p}_{\mathbf{k}-\mathbf{q}})^2}\right\}\label{Pi_x_nc}
\end{multline}
is given by the RPA-type one-loop polarization diagram with two virtual Bogoliubov quasiparticles. Here
\begin{equation}
\left\{\begin{array}{c}u_\mathbf{k}\\v_\mathbf{k}\end{array}\right\}=\pm\sqrt{\frac12\left(\frac{\tilde\epsilon^\mathrm{p}_k + X_0^2X_k^2n_0^\mathrm{p}\tilde{v}_k^\mathrm{xx}}{E^\mathrm{p}_{\mathbf{k}}}\pm1\right)}
\end{equation}
are the bosonic Bogoliubov coefficients for polaritons, and $f_{\mathbf{k}}=\{\exp(E^\mathrm{p}_k/T)-1\}^{-1}$ are the Bose-Einstein occupation numbers of Bogoliubov excitations with the energies $E^\mathrm{p}_k$, which, for the estimation purpose, can be calculated in TFA using (\ref{E_q_TFA}).

The integration over momenta $\mathbf{k}$ in (\ref{Pi_x_nc}) is logarithmically divergent at $|\mathbf{k}|\rightarrow0$ in two dimensions, which is related to the absence of a true Bose condensate in 2D systems \cite{Kagan,Mora}. To obtain a physically meaningful result we impose the long-wavelength cutoff $|\mathbf{k}|>2\pi/L$, where $L\sim10\,\mu\mbox{m}$ is the linear size of the polariton BEC cloud. According to our numerical calculations, in realistic conditions at any $(q,i\omega_n)$, $\Pi^\mathrm{nc}_\mathrm{x}$ is at least 3 orders of magnitude smaller than the leading-order condensate term $\Pi_\mathrm{x}$.

Therefore we conclude that the noncondensate processes are negligible for the pairing. However in \cite{Sun_2DMater,Sun_Thesis,Sun_NJP,Sun_PRR} it is argued that the noncondensate processes should provide dominant contribution to the coupling constant, because the competing condensate contribution (\ref{Pi_x}) is suppressed, $\Pi_\mathrm{x}(q,i\omega_n)\rightarrow0$, at low momenta $q\rightarrow0$ by the factor $\tilde\epsilon_q^\mathrm{p}\propto q^2$ (as interpreted in terms of destructive interference between the coherence factors $u_\mathbf{k}$, $v_\mathbf{k}$), while the noncondensate contribution is free of such suppression, $\Pi^\mathrm{nc}_\mathrm{x}(q,i\omega_n)\rightarrow\Pi^\mathrm{nc}_\mathrm{x}(0,i\omega_n)\neq0$ at $q\rightarrow0$.

In our approach, the noncondensate contribution to $\Pi_\mathrm{x}$ and hence to $\tilde\Pi_\mathrm{x}$ is dressed in Eq.~(\ref{V_2}) with the square of the screened electron-exciton interaction (\ref{v_ex_tilde}), and thus in TFA it acquires the factor $[1-v_q^\mathrm{ee}\Pi_\mathrm{e}]^{-2}\propto q^2$ in the limit $q\rightarrow0$. This factor, originating from the screening, was omitted in \cite{Sun_2DMater,Sun_NJP,Sun_PRR}. It suppresses the long-wavelength contributions of both $\Pi_\mathrm{x}$ and $\tilde\Pi_\mathrm{x}$ to the pairing interaction, so the difference between their long-wavelength behaviors becomes unimportant. Moreover, as we show in Sec.~\ref{Sec3}, the dominating contribution to the coupling constant is provided by the softened Bogoliubov excitations near the roton minimum at $q=q_0\approx1/2L$, and not by the long-wavelength modes with $q\rightarrow0$. This circumstances explain why the pair-bogolon processes do not provide any appreciable contribution to the pairing constant in our analysis, where the interaction screening and formation of the roton minimum are taken into account.

\bibliography{References}

\end{document}